\documentclass[sigconf]{acmart}
\usepackage[ruled,vlined, linesnumbered]{algorithm2e}

\renewcommand\footnotetextcopyrightpermission[1]{}

\usepackage{amsmath,amssymb,amsfonts}
\usepackage{subfigure}
\usepackage{textcomp}
\usepackage{xcolor}
\usepackage{graphicx}
\usepackage{multirow}
\usepackage{textcomp}
\usepackage{tikz}
\usepackage{xcolor}
\usepackage{hyperref}
\usepackage{xspace}
\usepackage{listings}
\usepackage{bbding}
\usepackage{wasysym}
\usepackage{caption}
\usepackage{enumitem,kantlipsum}
\usepackage{textcomp}
\usepackage{xcolor}
\usepackage{soul}
\usepackage{interval}
\usepackage{pgfplots}
\usepackage{pifont}

\newcommand{\cmark}{\ding{51}}%
\newcommand{\xmark}{\ding{55}}%
\usepackage[noend]{algpseudocode}

\newcommand{\ours}{\mbox{PhantomSound}\xspace}
\newcommand{\rev}[1]{{\color{black} #1}}

\algrenewcommand\algorithmicforall{\textbf{foreach}}
\algrenewcommand\algorithmicindent{.8em}
\def\BibTeX{{\rm B\kern-.05em{\sc i\kern-.025em b}\kern-.08em
    T\kern-.1667em\lower.7ex\hbox{E}\kern-.125emX}}
\AtBeginDocument{%
  \providecommand\BibTeX{{%
    \normalfont B\kern-0.5em{\scshape i\kern-0.25em b}\kern-0.8em\TeX}}}

\copyrightyear{2023}
\acmYear{2023}
\setcopyright{acmlicensed}\acmConference[RAID '23]{The 26th International Symposium on Research in Attacks, Intrusions and Defenses}{October 16--18, 2023}{Hong Kong, Hong Kong}
\acmBooktitle{The 26th International Symposium on Research in Attacks, Intrusions and Defenses (RAID '23), October 16--18, 2023, Hong Kong, Hong Kong}
\acmPrice{15.00}
\acmDOI{10.1145/3607199.3607240}
\acmISBN{979-8-4007-0765-0/23/10}

\begin{document}
\title{PhantomSound: Black-Box, Query-Efficient Audio Adversarial Attack via Split-Second Phoneme Injection}
\author{Hanqing Guo}
\email{guohanqi@msu.edu}
\affiliation{%
\institution{Michigan State University}
\city{East Lansing}
\state{Michigan}
\country{USA}
}

\author{Guangjing Wang}
\email{wanggu22@msu.edu}
\affiliation{%
\institution{Michigan State University}
\city{East Lansing}
\state{Michigan}
\country{USA}
}

\author{Yuanda Wang}
\email{wangy208@msu.edu}
\affiliation{%
\institution{Michigan State University}
\city{East Lansing}
\state{Michigan}
\country{USA}
}

\author{Bocheng Chen}
\email{chenboc1@msu.edu}
\affiliation{%
\institution{Michigan State University}
\city{East Lansing}
\state{Michigan}
\country{USA}
}

\author{Qiben Yan}
\email{qyan@msu.edu}
\affiliation{%
\institution{Michigan State University}
\city{East Lansing}
\state{Michigan}
\country{USA}
}

\author{Li Xiao}
\email{lxiao@cse.msu.edu}
\affiliation{%
\institution{Michigan State University}
\city{East Lansing}
\state{Michigan}
\country{USA}
}
\begin{abstract}
In this paper, we propose \ours, a query-efficient black-box attack toward voice assistants.
Existing black-box adversarial attacks on voice assistants either apply substitution models or leverage the intermediate model output to estimate the gradients for crafting adversarial audio samples. However, these attack approaches require a significant amount of queries with a lengthy training stage. \ours leverages the decision-based attack to produce effective adversarial audios, and reduces the number of queries by optimizing the gradient estimation. In the experiments, we perform our attack against 4 different speech-to-text APIs under 3 real-world scenarios to demonstrate the real-time attack impact. The results show that \ours is practical and robust in attacking 5 popular commercial voice controllable devices over the air, and is able to bypass 3 liveness detection mechanisms with $>95\%$ success rate. The benchmark result shows that \ours can generate adversarial examples and launch the attack in a few minutes. We significantly enhance the query efficiency and reduce the cost of a successful untargeted and targeted adversarial attack by 93.1\% and 65.5\% compared with the state-of-the-art black-box attacks, using merely $\sim$300 queries ($\sim$5 minutes) and $\sim$1,500 queries ($\sim$25 minutes), respectively. 
\end{abstract}


%


\maketitle

\section{Introduction}\label{sec-introduction}
Voice
is the primary method for human-computer interaction. Driven by the unprecedented amount of voice data and flourishing development of Artificial Intelligence (AI) technology, the modern deep-learning based Automatic Speech Recognition (ASR) systems and Intelligent Voice Control (IVC) devices have been integrated into our daily lives. 
According to Voicebot's 2019 consumer report~\cite{kinsella2019smart}, it was reported that 26\% of U.S. adults own a smart speaker. 
Nowadays, users can directly speak to their smartphones to interact with the voice assistants such as Siri~\cite{siri}, Google Assistant~\cite{GoogleAssistant}, or smart speaker systems such as Google Home~\cite{GoogleHome}, Amazon Echo~\cite{Echo}. 

The voice commands have been used to send and read text messages, make phone calls, set timers, check calendar entries, and even order a drink from Starbucks or summon a Uber with ``skills"~\cite{skills}. 
More and more tech companies now provide ASR services, including Amazon Transcribe~\cite{amazon_transcribe}, Google Cloud Speech-to-Text~\cite{Google_speech}, IBM Watson Speech to Text~\cite{IBM_Speech}, and Microsoft Azure Speech Service~\cite{microsoft_azure}, all of which allow the developers to empower their apps with intelligent audio functionalities. 

However, with the increasing presence of ASR systems and IVC devices in private spaces, users begin to worry about the security and privacy of these systems. 
For example, a hacked device is now capable of recording private conversations;
collecting and sharing private data; and controlling all the connected IoT devices in smart homes~\cite{chen2020devil, sugawara2020light}. Researchers have demonstrated that ASR systems could become vulnerable to a wide variety of attacks. For instance, inaudible commands can be injected through ultrasound~\cite{zhang2017dolphinattack, roy2018inaudible},
even across different transmission media, such as object surface~\cite{yan2020surfingattack}, light~\cite{sugawara2020light}, etc. Besides the physical attacks, recent studies also utilize the discrepancies between the human ear and feature extraction algorithms to launch \emph{signal processing attacks}~\cite{abdullah2019practical, abdullah2021hear}. 

Despite the aggravating threats, these new attacks could be defeated by integrating additional hardware~\cite{zhangeararray} or extra signal processing procedures (e.g., voice activity detection, guard signals)~\cite{abdullah2019practical, he2019canceling}. 
Unlike the aforementioned attacks, the \emph{adversarial attack} aims to attack the deep neural networks (DNN), i.e., the computational core of an ASR system, which poses a major threat to modern ASR systems.

\vspace{5pt}
\noindent \textbf{Adversarial Attack:} Adversarial attack was first proposed to attack image recognition systems~\cite{goodfellow2014explaining, szegedy2013intriguing}.  
The attack operates by imposing unnoticeable perturbations onto the original image, thereby misleading the DNN to yield false classification. The inputs that enable such an attack are commonly referred to as \emph{Adversarial Examples (AEs)}, which are composed of the original input with an unnoticeable perturbation. The ASR system with DNN models also inherits the susceptibility 
towards AEs. 

\vspace{5pt}
\noindent\textbf{Prior Studies:}
Prior studies~\cite{cisse2017houdini, carlini2018audio, alzantot2018did} demonstrate that attackers can generate adversarial audios to alter the DNN's prediction result with or without the prior knowledge of the DNN model. However, most of these attacks have not been successfully realized against real-world commercial devices, and their stealthiness is unverified. Recently, Chen et al.~\cite{chen2021real} successfully attack both open-source and commercial speaker verification systems over the air in a grey-box setting. 
Yuan et al.~\cite{yuan2018commandersong} embed their generated AE within songs to launch the attack, and they further adapt their attack in a black-box setting to subvert the ASR of most IVC devices~\cite{chen2020devil}. Nevertheless, they fail to guarantee the attack success rate in the presence of user interference; and cannot promise to craft AEs quickly due to the training overhead of the substitution model. 
Meanwhile, two recent studies~\cite{li2020advpulse, guo2022specpatch} inventively propose the sub-second perturbation and spectrogram patch perturbation to attack open-source ASR systems, considering the victim user present during the attack. Even though they demonstrate the robustness and feasibility of their attack 
in the presence of environmental distortions, the proposed attacks are established on the assumption of complete knowledge of the target ASR system. More recently, Zheng \textit{et al.} propose a decision-based black-box attack by incorporating evolutionary algorithms to generate adversarial audios~\cite{zheng2021black}. 
However, they still require to query the victim model extensively, which incurs substantial time and financial costs in a practical attack scenario. 

\begin{figure}[t]
    \centering
    \includegraphics[width=0.7\linewidth]{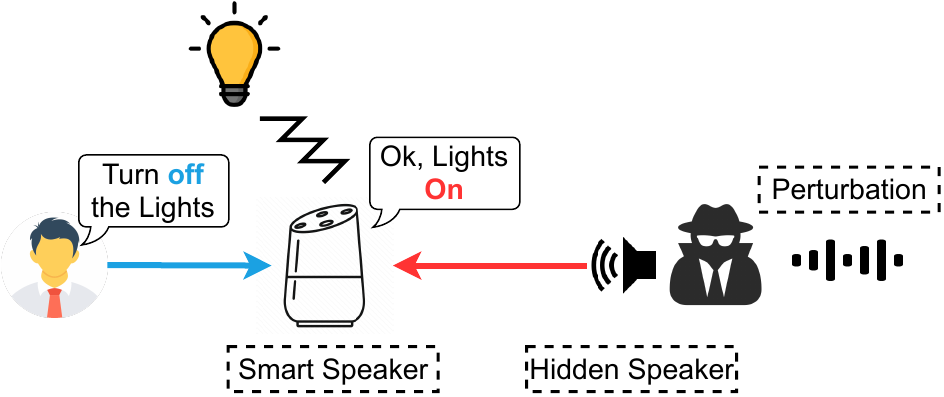}
    \caption{Attack scenario of \ours}
    \label{fig:cover}
\end{figure}
 Table~\ref{tab:comparison1} summarizes the existing adversarial attacks in terms of \emph{victim systems' tasks}, \emph{attacker knowledge}, \emph{ability to attack quickly}, and \emph{attack scenario}. 
 \rev{The check mark symbolizes a successful attack under the given scenario, while the cross mark implies that the attack could not function or lacks efficacy in that particular scenario.} For the \emph{victim system's task}, SV indicates the speaker verification task while 
 SR refers to the speech recognition task. We then taxonomize \emph{attacker knowledge} into white-box, grey-box, and black-box, where grey-box implies the attacker can get the logits layer output~\cite{alzantot2018did, chen2021real} or confidence score of all possible classes, and black-box indicates the attacker can only access the prediction label~\cite{chen2020devil} of the target model. 
 A white-box attacker, on the other hand, has complete knowledge (model architecture, weights of DNN parameters) of the target system. Next, we use online AE generation (Online GENR) to characterize whether the attacker can generate AEs or perturbations swiftly and complete the attack procedure in an online fashion. In fact, most existing studies assume the attacker has sufficient time to produce AEs offline. The last two metrics, Over Air and User Interference (User INT) suggest the attack scenario, where the former indicates an over-the-air attack, while the latter indicates whether the attack considers the user's interference (e.g., voice commands) during the attacks. 
 To the best of our knowledge, \emph{no existing attacks can attack commercial, closed-source ASR systems over-the-air with a limited time budget and user interference.}
\begin{table}
    \caption{Comparison with other recent audio attacks.}
    \label{tab:comparison1}
    \centering
    \scalebox{0.8}{
    
        \begin{tabular}{p{2.2cm}p{0.5cm}p{0.6cm}p{0.6cm}p{0.7cm}p{0.6cm}p{0.6cm}}
        \hline
          \multirow{2}{*}{\small{\textbf{Attacks}}} & \small{\textbf{SV SR}} &
          \small{\textbf{Grey Box}} &
          \small{\textbf{Black Box}} &
          \small{\textbf{Online GENR}} & \small{\textbf{Over Air}} & \small{\textbf{User INT}}\\
          \hline
          Houdini~\cite{cisse2017houdini} & SR  & \cmark & \xmark & \xmark & \xmark & \xmark \\
          \hline
          C\&W~\cite{carlini2018audio} & SR  & \xmark & \xmark &\xmark & \xmark & \xmark \\
          \hline
          Adversarial~\cite{alzantot2018did} & SR  & \cmark & \xmark &\xmark & \xmark & \xmark \\
          \hline
          Fakebob~\cite{chen2021real} & SV  & \cmark & \xmark & \xmark & \cmark & \xmark \\
          \hline
          Comm.~\cite{yuan2018commandersong} & SR  & \xmark & \xmark & \xmark & \cmark & \xmark \\
          \hline
          Devil's~\cite{chen2020devil} & SR  & \cmark & \cmark & \xmark & \cmark & \xmark \\
          \hline
          AdvPulse~\cite{li2020advpulse} & SR  & \xmark & \xmark & \xmark & \cmark & \cmark \\
          \hline
          OCCAM~\cite{zheng2021black} & SR  & \cmark & \cmark & \xmark & \cmark & \xmark \\
          \hline
          SpecPatch~\cite{guo2022specpatch} & SR  & \xmark & \xmark & \xmark & \cmark & \cmark \\
          \hline
          \hline
          \textbf{\ours} & \textbf{SR}  & \cmark & \textbf{\cmark} & {\cmark} & {\cmark} & {\cmark} \\
          \hline
      \multicolumn{5}{l}{
    \vspace{-20pt}} \\
        \end{tabular}
        }
\end{table}

\vspace{5pt}
\noindent \textbf{\ours:} We propose a query-efficient black-box attack on commercial closed-source ASR systems and IVC devices. Our attack, called \emph{\ours}, can craft AEs and perturbations within a limited time budget and restricted query cost. 
Different from the previous work, the key idea behind \ours is to regard the users' voice input as the command ``carrier", while the phoneme-level perturbations are applied on the ``carrier" to instantiate the attack. 

Figure~\ref{fig:cover} depicts the attack scenario. First, the adversary records the user's command (any keywords such as ``open", ``on", ``down"). Next, the adversary uses \ours to query the accessible target models on the target IVC devices (e.g., the Google Cloud Speech-to-Text API for Google Home). Then, \ours returns a perturbation that alters the prediction of the user's command. 

During the attack, the adversary plays the perturbation via a hidden speaker at the same time when the user utters a voice command, which fools the smart speaker to operate improperly.

\vspace{5pt}
\noindent \textbf{Challenges:} Four major challenges arise during the design of \ours. 
\begin{itemize}[leftmargin=*]
    \item \textbf{Black-box Attack:} It is difficult to attack a model without any prior knowledge.
    Existing grey-box/black-box attacks either assume attackers have the probability score of the target model~\cite{cisse2017houdini, alzantot2018did}, or train a substitution model to approach the target model~\cite{chen2020devil}. The existing attacks require a substantial amount of time to train a substitution model for the generation of AEs. 
    \item \textbf{Speech Model:} Different from black-box attacks on image processing~\cite{chen2020hopskipjumpattack, cheng2019sign}, ASR systems are known to have a more complicated model structure consisting of signal processing, filtering, acoustic model, and language model. As a result, attacking speech models requires different attack strategies to bypass the various components of the ASR models. 
    \item \textbf{Query Efficiency:} 
    A successful black-box attack relies excessively on the effectiveness of queries. The adversary needs to iteratively update the AEs such that the effectiveness of the crafted AEs can be justified through querying. However, querying commercial ASR APIs is costly (e.g., \$0.00001/second for Google Cloud Speech-to-Text) and unable to bypass. Despite some efforts~\cite{chen2020hopskipjumpattack, cheng2019sign} to reduce the number of queries, it still  falls short of meeting the requirements  for online generation of AEs.
    \item \textbf{Perturbation Sync:} To successfully launch our attack, the adversary is expected to play the perturbation when he/she hears the victim's voice command. However, in a real-world scenario, the timing of perturbation is hard to control. Therefore, we need to tackle this problem by generating a near-synchronization-free
    perturbation~\cite{li2020advpulse}. 

\end{itemize}

\vspace{5pt}
\noindent \textbf{Contributions:}
The contributions of this work are highlighted as follows.
\begin{itemize}[leftmargin=*]
    \item \textbf{New Attack:} To the best of our knowledge, we are the first to achieve query-efficient black-box attacks on commercial ASR systems as well as IVC devices. We demonstrate the dangers of our attack over-the-air  on 4 different commercial ASR APIs (i.e., Google Cloud Speech-to-Text, IBM Watson Speech to Text, Amazon Transcribe, and Microsoft Azure Speech Service) and 5 different IVC devices (i.e., iPhone with Google Assistant, Google Home, Microsoft device, Amazon Alexa, and IBM Wav-Air-API).
    \item \textbf{New Finding:} We discover and formulate the unique boundary of commercial ASR systems for producing AEs. This non-contiguous decision boundary hinders previously successful attempts.

    \item \textbf{New Techniques:} We propose \ours, a phoneme-level searching method for efficiently crafting AEs to launch adversarial perturbation attack  with the least number of required queries in comparison with other methods.
\end{itemize}



The remainder of the paper is organized as follows. We introduce the background and preliminary observations in $\S$\ref{sec-background}, followed by the system design of \ours in $\S$\ref{sec-system}. 
The implementation and evaluation are further presented in $\S$\ref{sec-evaluation}. The discussion and limitation of \ours are entailed in $\S$\ref{sec-discussion-and-limitation}. We present the related work in $\S$\ref{sec-relate} and conclude the paper in $\S$\ref{sec-conclusion}.

\section{Background and Preliminary Study}\label{sec-background}

\begin{figure*}[t!]
\centering     
\subfigure[A mixed image with cat and dog is recognized by Google Cloud Vision API~\cite{Google_Vision} with 89\% cat and 11\% dog.
]{\label{fig:a}\includegraphics[width=0.23\linewidth]{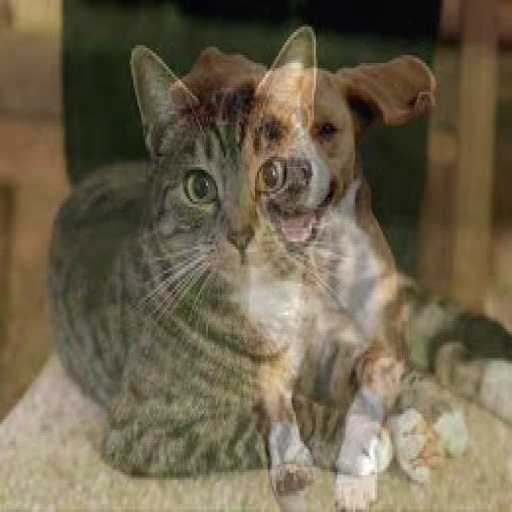}}
\subfigure[Search decision boundary in black-box CV attack.]{\label{fig:c}\includegraphics[width=0.23\linewidth]{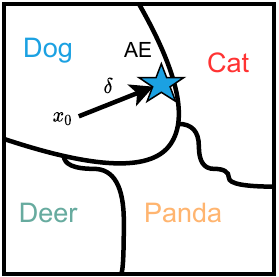}}
\subfigure[A mixed audio with ``stop" and ``backward" is rejected by
Google Speech-to-Text API with no output]{\label{fig:b}\includegraphics[width=0.23\linewidth]{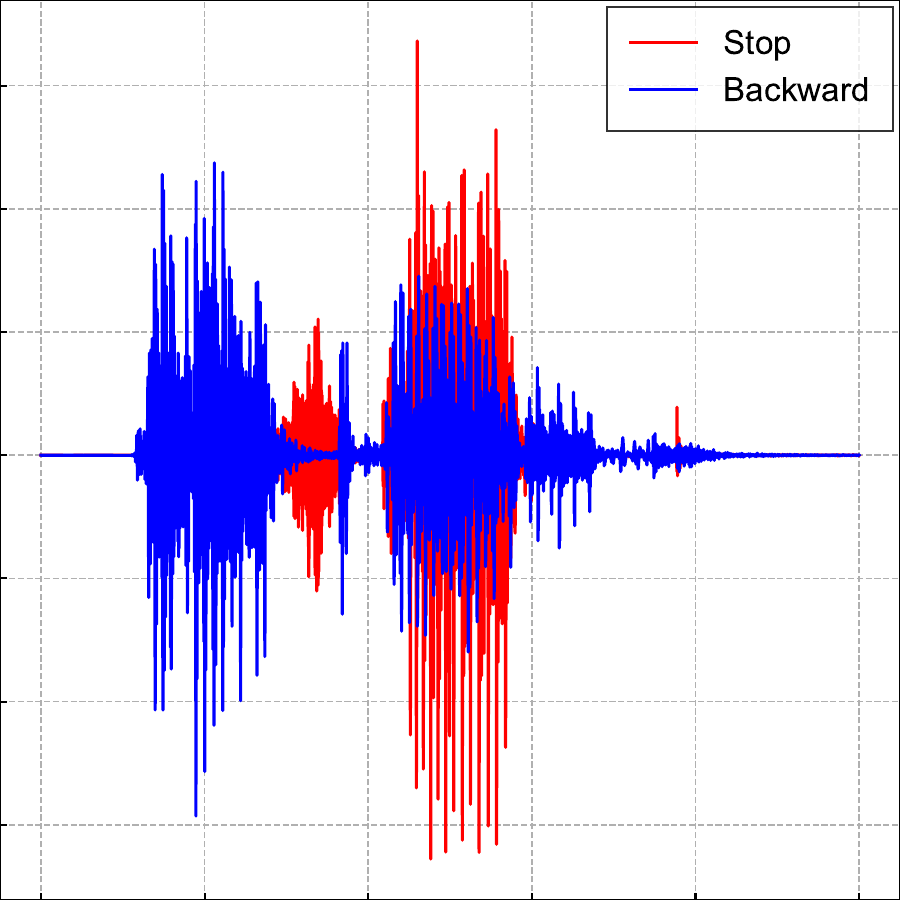}}
\subfigure[The decision boundary for every class is non-contiguous for ASR system, every input in the middle will be rejected due to ambiguity.]{\label{fig:d}\includegraphics[width=0.23\linewidth]{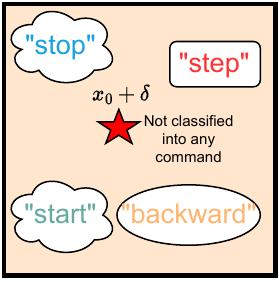}}
\vspace{-10pt}
\caption{Observations of CV and ASR systems.}
\end{figure*}

In this section, we present the threat model of \ours, as well as the assumptions and attack scenarios. Then, we introduce the fundamentals behind the adversarial attack and present the decision scheme of commercial ASR systems. 

\subsection{Threat Model}
\rev{The adversary's goal is to mislead the IVC devices or VCS systems by injecting malicious commands. Prior to our work, there are two types of attacks that can achieve the same goal. The first attack~\cite{chen2020devil} uses reverse-engineering models to imitate the commercial models and craft the offline AE in a white-box manner. The second attack~\cite{wenger2021hello} uses generative models to synthesize the victim's speech. 
However, the reverse-engineering attack necessitates a high volume of queries (as per Table~\ref{tb:cost}) to construct the substitute model. It also demands updating the model in response to changes in the commercial API. This renders it expensive and inadequate in meeting the need for a real-time attack. Regarding the generative model driven synthesized attack, we assume the adversary has access to sufficient recordings of the victim for training purposes. However, in our specific situation, the attacker is expected to initiate the attack upon their first encounter with the victim. Furthermore, playing the synthesized speech outright is not a viable approach as the victim can hear it and potentially halt the attack.
}

\noindent\textbf{Adversary's Capability:} We assume that the adversaries can place a hidden microphone to record the victim's voice. We assume that an adversary knows the targeted IVC devices and has access to their respective ASR API services (e.g., Google Cloud Speech-to-Text for Google Home or Google Assistant). Following other related studies~\cite{yuan2018commandersong, chen2020devil, zheng2021black, li2020advpulse, alzantot2018did}, we also assume that the adversary is able to launch this attack via a hidden speaker or a compromised speaker in the victim's workspace/home.

\noindent\textbf{Attack Scenarios:} 
The adversary will first collect the victim's voice commands, and then generate the AEs and perturbations swiftly only based on the transcription result of the target devices. Once the perturbations are crafted, the adversary can wait for the victim's next command and play the perturbation manually or automatically via existing keyword searching or voice detection mechanisms~\cite{tang2018deep, alvarez2019end}. 
Alternatively, the adversary may also play the perturbation repeatedly through hacked speakers, attempting to fool the target IVC devices when the corresponding target voice command was delivered. 

In a real-world attack scenario, e.g., in a public space, an attacker may not have access to a large collection of victims' voices and may not have sufficient time to generate the perturbation offline. 
In this case, the attacker only has a very limited time window to subvert the victims' 
commands towards voice assistants. 
To successfully instantiate such an opportunistic voice attack, an attack approach with a timely and low complexity AE generation is highly desired.

\noindent\textbf{User Interference:}
Most existing attacks assume that the users will not perceive the AEs and will not interact with their voice assistants during the attack. However,  when the users are speaking during the attack, most existing voice attacks will fail.
In this research, we leverage the users' voice command as a carrier for the adversarial audio to launch the attacks more effectively and stealthily. 
Moreover, as advanced liveness detection algorithms~\cite{ahmedvoid, li2021robust} have been used to differentiate between loudspeakers and humans with high accuracy, most existing audio attacks launched by loudspeakers can be easily detected. 
In our attack, however, since the human voice and the perturbation arrive at the same time, the liveness detection module of the voice assistant can be effectively bypassed.

\subsection{Adversarial Attack}
Adversarial attack aims to craft an AE  $x_0+\delta$, in order to deceive the model $f(\cdot)$ to make false prediction~\cite{szegedy2013intriguing}. 
Take $y_{pred}$ as the output of model, 
if $f(x_0+\delta):=y_{pred}\neq y$ ($y$ indicates the true label of input $x_0$), we suppose the attacker has launched an untargeted attack. If the perturbation is crafted intentionally for a specific target (denoted as $y_t$), the attack formalized as $f(x_0+\delta)=y_{t}\neq y$, is regarded as a targeted attack. 
The generation of AE can be formulated as an optimization problem as follows: 
\begin{equation}
    minimize\quad \mathcal{L}(x_0+\delta) := \mathcal{D}(f(x_0+\delta), y_t). 
    \label{eq:loss}
\end{equation}
The goal of Eq.~(\ref{eq:loss}) is to minimize $\mathcal{L}(x_0+\delta)$ under the constraint that $||\delta||_2<\epsilon$, where $\mathcal{L(\cdot)}$ denotes the loss function, which uses a distance function $\mathcal{D(\cdot)}$ to measure the disparity between $f(x_0+\delta)$ and $y_t$, $||\cdot||_2$ is the L2 norm, and $\epsilon$ is used to control the amplitude of perturbation. There are three main types of attacks depending on the prior knowledge of the victim models, listed as follows:  

\noindent\textbf{White-box:} If the adversaries learn architecture and the parameters of the model, they can get the gradient of the loss function $\nabla \mathcal{L}(x)$ during the forward or backpropagation. The perturbation can be subsequently estimated using the inverse gradient~\cite{goodfellow2014explaining}. 

\noindent\textbf{Grey-box:} The model conceals its architecture and parameters from the public and only exposes the prediction scores $P = [p_0, p_1,$ 
$\cdots, p_n]$ for a given input. The adversaries can formulate a loss function~\cite{carlini2017towards} $\mathcal{D}(P, P_y)$ ($P_y$ is the one-hot encoding of $y$), and then track the changes of distance when tuning $\delta$ in multiple attempts. The changes in $\mathcal{L}(x)$ are utilized to estimate the gradient which will guide the attacker to update $\delta$. The gradient estimation algorithms include Natural Evolution Strategy (NES)~\cite{ilyas2018black} and Zeroth Order Optimization (ZOO)~\cite{chen2017zoo}.

\noindent\textbf{Black-box:} Compared to white-box and grey-box attacks, the black-box attack is the most challenging, in which the attacker only has access to the prediction label of the model. In fact, most of the commercial ASR systems and IVC devices are closed-source and only offer a final prediction. To successfully attack the black-box model, existing work either trains a surrogate model and transforms the problem into a white-box attack~\cite{papernot2017practical}, or uses a significant amount of queries to search the decision boundary of the victim model~\cite{brendel2017decision, cheng2019sign, chen2020hopskipjumpattack}. 
Here, we focus on the query-based boundary-searching attack due to its  flexibility and attack efficiency.

\subsection{Black-box Audio Adversarial Attack}
Compared with the black-box adversarial attack in other domains, the black-box audio adversarial attack has several unique features. 
In this section, we conduct a preliminary study in quantifying the behaviors of commercial ASR services.

\noindent\textbf{Decision-based Attack:} Used for classification, a decision boundary is a hypersurface that partitions the sample space into several classes. 
Specifically, a well-trained DNN model uses the decision boundary to classify the incoming inputs. The main goal of the existing black-box attacks~\cite{brendel2017decision, cheng2019sign, chen2020hopskipjumpattack}, or so-called decision-based attacks, is to find the decision boundary of the target model. Generally, to approach the precise decision boundary, they gradually perturb the input based on the query feedback, to find an AE on the verge of the decision boundary. 

However, one assumption made by existing decision-based attacks is that the DNN classification model guarantees to return a prediction $y_{pred}$ for any input $x$. As shown in Fig.~\ref{fig:a}, we merge a cat and a dog into one image and feed it into Google Cloud Vision API~\cite{Google_Vision}. The classifier labels the image as a cat with very high confidence (89\%) while the human brain perceives it differently. 
As shown in Fig.~\ref{fig:c},
the decision-based adversary~\cite{chen2020hopskipjumpattack} starts from a dog ($x_0$) and adds the proportion of a cat ($\delta$) gradually to approach the boundary. The curves between classes in Fig.~\ref{fig:c} indicate the decision boundaries, where $\delta\in [0, 255]^{H\times W}$ denotes the perturbed image with the same shape as $x_0$. The contiguous decision boundary allows the DNN models to always output a result, while the result turns unreliable as it approaches the decision boundary.

\noindent\textbf{Decision Boundary of ASR:} At first sight, it appears that the ASR systems would inherit the DNN's susceptibility to decision-based adversarial attacks. 
However, the unique characteristics of voice systems and DNN models make traditional decision-based attacks hard to succeed. 
Here, we conduct a preliminary experiment, in which we mix two voice commands ``stop" and ``backward" together (Fig.~\ref{fig:b}) to imitate the mixture of cat and dog images. Then, we submitted the mixed audio to Google Speech-to-Text API, which was rejected without any returns. The failed attempts indicate that the decision boundary of the ASR system is non-contiguous. As shown in Fig.~\ref{fig:d}, every voice command is surrounded by an exclusive boundary, and the audios outside of the boundary ranges will be rejected by the ASR systems. 

This phenomenon implies that the perturbed voice queries may fail to solicit valid feedback from the ASR systems. Without feedback, it is difficult to determine the direction of the perturbation for approaching a target decision boundary. 
Based on this observation, we are motivated to design a new boundary-searching method to enable the decision-based black-box attack toward ASR systems.

\section{Attack Design}\label{sec-system}
In this section, we present the system design of \ours. We first introduce 
the phoneme-level boundary searching method to minimize the possibility of rejection by the ASR systems. Then, we formalize the attack as an optimization problem and illustrate the generation of AEs.  
Finally, to enhance the robustness of \ours in real-world scenarios, we propose the weak synchronization scheme and over-the-air speech enhancement.

\subsection{Phoneme-level Boundary Searching}
Fig.~\ref{fig:d} shows the challenge in boundary searching to produce a proper AE. If the adversary randomly adds noise to ``stop", the ASR remains the ``stop" decision when the noise is low and gives rejection while rising the noise power. However, if the adversary directly applies target ``backward" to the benign audio, it results in audio (red start in Fig.~\ref{fig:d}) in the middle between two decision boundaries, hence giving no output.

Therefore, the reasons behind the rejection of queries can be attributed to two factors: 1) the added random noise will elevate the command's noise level; 2) the boundary distance between two valid commands is too long to allow for an unnoticeable perturbation. 
\begin{figure}[t]
    \centering
    \includegraphics[width=0.8\linewidth]{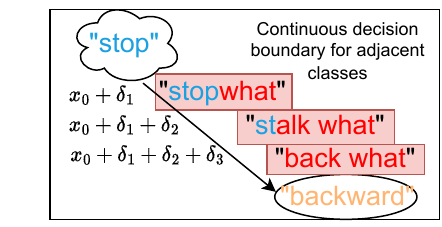}
    \caption{Phoneme guided query}
    \label{fig:ba2}
\end{figure}
To resolve these two problems, a novel idea is raised: ``If we break the target ``backward" into small pieces, then craft AE with sub-targets which directly connect to the benign decision boundary with small pieces, 
and finally, we can craft the final AE with the target." Fig.~\ref{fig:ba2} depicts our attack design. Specifically, instead of directly adding ``backward" on the ``stop",
we break the target ``backward" into a series of phonemes. During crafting the AE, we randomly add the phoneme on the benign audio and check the prediction. If the ASR produces a word that is closer to our target, we keep the phoneme on the benign audio and search for a closer prediction in the next round. In our case, the ``stop" adds perturbation phoneme $\delta_1$ and is recognized as ``stopwhat", then changes to ``stalk what", and ``back what", and finally reaches the target ``backward". In every step, the AE achieves to sub-targets who is adjacent to the benign decision boundary, and gradually, the perturbation can be crafted by summing up all the small changes.



The basic idea of the proposed phoneme-level searching method is to perturb the original command along the direction of the target command while minimizing the distance between the original and the target ones.


Algorithm~\ref{al:init} presents the initialization procedure for generating the phoneme-level adversarial perturbation. Specifically, we first set the counter $s=0$, and the initial distance between benign and target as $\epsilon$ = CER($f(x_0), y_t$). Next,
we construct a phoneme set $D=\{ph_1, ph_2, ..., ph_n\}$ by breaking the target command, and then generate a random noise $v \in [0, 0.1]^l$ in line 4, where $l$ is the length of original input $x_0$. Next, together with the $v$, a phoneme from $D$ is randomly picked and injected at its corresponding position of $x_0$ in lines 5-6 to generate an AE $x_*$. The purpose of $v$ is to increase the variance of the phoneme.
For the targeted attack, if the $x_*$ has a smaller distance to the target (line 7), we put the perturbation to the initial perturbation set $\Tilde{P}$, then update the $\epsilon$ and $x_0$. For an untargeted attack, we can replace line 7 with ``if $f(x^*)!=y$" to assure the ASR gives an incorrect prediction.
The searching loop continues until it reaches a sufficient number of rounds $K$.

\begin{algorithm}
\caption{Phoneme-level Adversarial Perturbation Initialization}
\label{al:init}
\SetAlgoLined
\SetKwInput{KwInput}{Input}
\SetKwInput{KwOutput}{Output}
\KwInput{The original audio $x_0$, the target label $y_t$, the phoneme clip samples $D=\{ph_1, ph_2, ..., ph_n\}$, the initial Character Error Rate(CER) $\epsilon$, the API service of black-box ASR system $f(\cdot)$.}
\KwResult{The initial perturbations set $\Tilde{P}$ }
 s = 0\;
$\epsilon$ = CER($f(x_0), y_t$)\;
 \While{$s < K$}{
  $v$ = random $[0, 0.1]^l$\;
  $\delta = v + rand(D) $\;
  $x^* = x_0 + \delta$\;
  \eIf{CER($f(x^*), y_t$) < $\epsilon$}{     
   Put $\delta$ into $\Tilde{P}$\;
   $\epsilon$ = CER($f(x^*), y_t$)\;
   $x_0 = x_0 + \delta$\;
   }{
   $s = s + 1$\;
  }
 }
 return $\Tilde{P}$
\end{algorithm}

\vspace{-10pt}
\subsection{Perturbation Optimization}
Even though Algorithm~\ref{al:init} generates proper perturbations for any voice commands, the amplitude of the perturbation may become overwhelming. Revisiting Eq.~(\ref{eq:loss}), to acquire the minimal perturbations, we need to gradually increase the perturbation power. However, due to the black-box setting, the gradient is inaccessible. As a result, we use Sign-Opt~\cite{cheng2019sign} to estimate the gradient, since Sign-Opt has achieved superior performance with the least number of queries, as written below: 

\begin{equation}
    \nabla \mathcal{L}(x) \approx \sum_{q=1}^{Q} sign(\mathcal{L}(x+\sigma\mu_q) -\mathcal{L}(x))\mu_q, 
    \label{eq:sign-opt}
\end{equation}

\begin{equation}
    sign(\mathcal{L}(x+\sigma\mu) -\mathcal{L}(x)) = 
    \begin{cases}
  +1, & f(x+\sigma\mu) \neq y_t \\
  -1 & f(x+\sigma\mu) = y_t
\end{cases}
    \label{eq:sign}
\end{equation}
where $x$ is the general representation of $x_0+\delta$, $q$ and $Q$ denote the noise index and the total number of noises respectively. $\sigma$ is the search variance and $\mu$ is the noise. 
The key idea of Sign-Opt is to search the gradient space using the natural evolution strategy. 
Since $\mathcal{L}(x)$ is unknown, Sign-Opt queries $f(\cdot)$ in Eq.~(\ref{eq:sign}). The feedback of the target model can be collected to measure the number of wrong predictions. The result will be used to guide Eq.~(\ref{eq:sign-opt}) in searching for the gradient of $\mathcal{L}(x)$.

\noindent\textbf{Query-Efficient Fine-tuning:} 
The perturbation generation typically requires $\sim$5k queries to craft an AE~\cite{cheng2019sign, chen2020hopskipjumpattack}. To further reduce the cost of queries, we design a query-efficient AE generation scheme to greatly reduce the query number. 


By carefully examining the Eq.~(\ref{eq:sign-opt}), we realize that the gradient estimation step depletes most of the queries. 
Suppose $Q=50$, then it uses $50$ queries to catch the $f(\cdot)$ result and estimate gradient according to Eq.~(\ref{eq:sign}). However, Sign-Opt~\cite{cheng2019sign} uses the estimated gradient only once for updating $x$, with a small update learning rate, while most of the gradient computations are wasted. In our design, we estimate the gradient once, then apply the estimated gradient multiple times to update the $\delta$ until it does not satisfy our attack goal, then do the gradient estimation again.



\begin{figure*}[t!]
    \centering
    \includegraphics[width=0.8\linewidth]{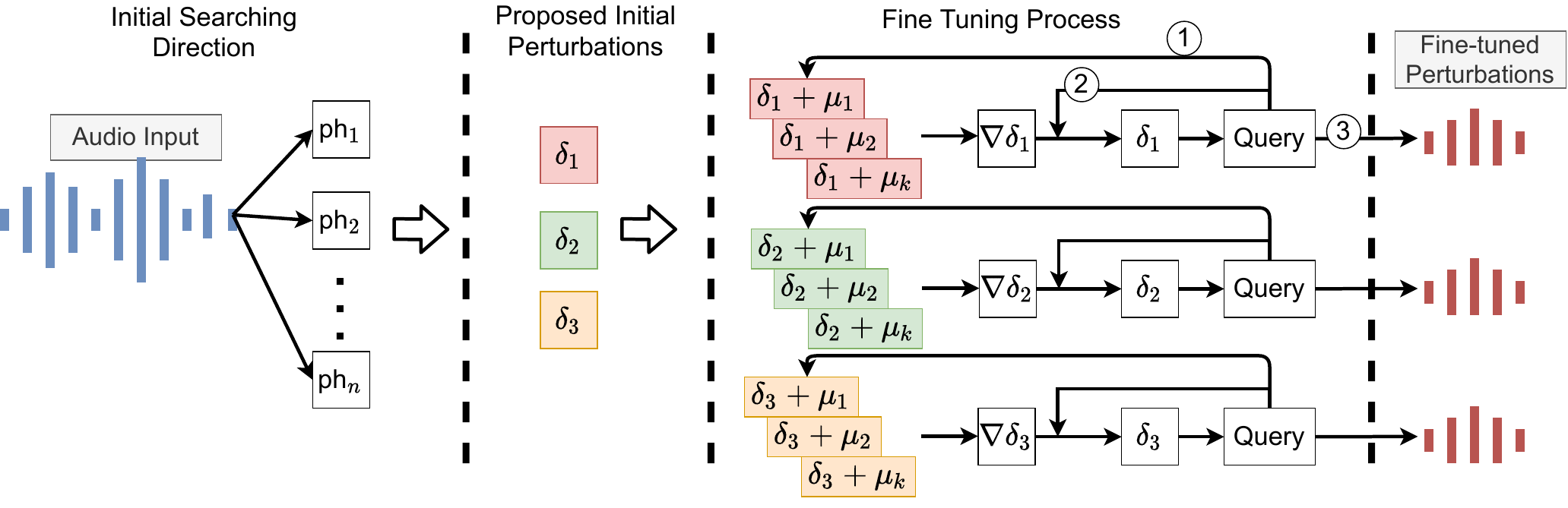}
    \vspace{-5pt}
    \caption{Adversarial perturbation generation}
    \label{fig:system}
    \vspace{-5pt}
\end{figure*}

\begin{figure}[h]
    \centering
    \includegraphics[width=0.7\linewidth]{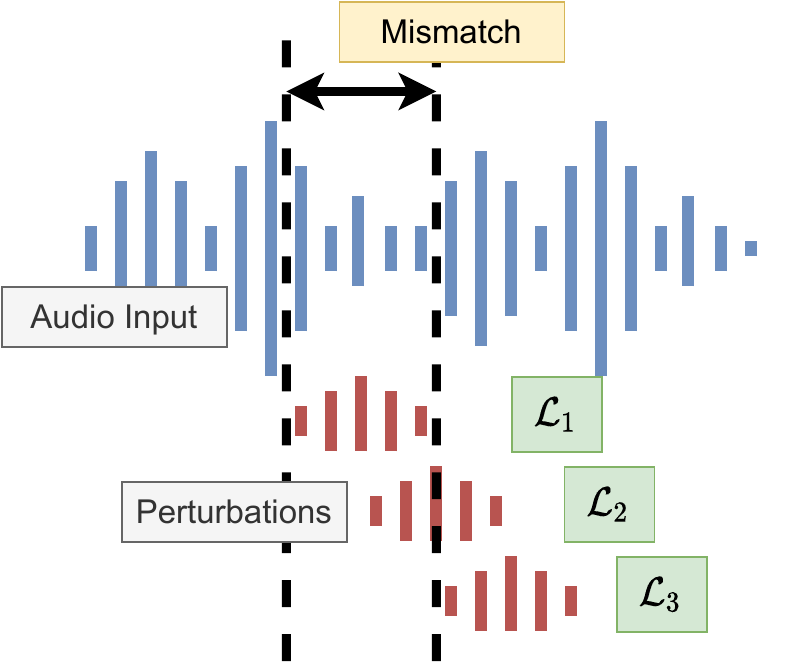}
    \caption{Perturbation mismatch during an attack. }
    \label{fig:mismatch}
\end{figure}

The workflow of our proposed query efficient phoneme-level adversarial perturbation generation is shown in Fig.~\ref{fig:system}.
There are three major steps to generate AEs and perturbations:  \emph{searching, proposing}, and \emph{fine-tuning}. In the searching and proposing phases, unlike the prior study~\cite{cheng2019sign} which only searches for random noise and keeps the shortest initial perturbation while discarding others, we reserve all the perturbation candidates to increase the generation speed. In the fine-tuning phase, we optimize all the proposed perturbations through gradient estimation. Note that there are three paths from the Query block: \textcircled{\small{2}} is used to update the perturbation consecutively until it cannot be further optimized. Then,  we will  re-calculate the gradient (\textcircled{\small{1}}). Once the power of perturbation is lower than $\epsilon$, we add it into the perturbation set $P$ (\textcircled{\small{3}}).

\subsection{Weak Synchronization Design}
\label{sec:weak_sync}
Considering the adversary needs reaction time to play the perturbation,
the generated perturbations are demanded to be robust against the mismatch of insertion positions. To realize such an attack, we seek to minimize the average loss instead of the instant loss. That is, we take the impact of mismatch into consideration and expect the comprehensive loss to be minimized.
Mathematically, the average loss can be expressed as follows:
\begin{equation}
    \overline{\mathcal{L}(x)} = \frac{1}{N}\sum_{i=1}^{N} \mathcal{L}_i(x),
    \label{eq:sync_1}
\end{equation}
\begin{equation}
    \mathcal{L}_i(x) = \mathcal{L}(x+c\tau),
    \label{eq:sync_2}
\end{equation}
where $\tau$ represents the mismatch interval, $c$ controls the length of a mismatch period, $i$ indicates the id of related losses, and $N$ is the number of involved $\mathcal{L}$. To minimize the average loss, we can refer to Eq.~(\ref{eq:sign-opt}) and Eq.~(\ref{eq:sign}) to estimate $\nabla \overline{\mathcal{L}(x)}$ by computing $\nabla\mathcal{L}(x+c\tau)$. The drawback of the average loss gradient estimation is that it costs $N\times $ more queries to perform the gradient estimation. 
The length of phonemes in $D$ varies from $50ms$ to $300ms$, and one-word duration is ranging from $281ms$ to $387ms$ according to the report~\cite{trauzettel2012standardized}. We expect that the phoneme-level perturbation can be plugged within the duration of one word, otherwise, it will be difficult to maintain the minimal $\mathcal{L}$ especially when a delayed perturbation arrives. In this paper, we set the $N=4$ and $\tau=100ms$. Fig.~\ref{fig:mismatch} depicts the perturbation mismatch scenario: when crafting the first red perturbation, we gather the other losses  by the same perturbation but with a different time delay. In the figure, $\mathcal{L}_1$, $\mathcal{L}_2$, $\mathcal{L}_3$ correspond to $c=0$, $c=1$, and $c=2$.  


\subsection{Over the Air Attack Robustness}
Besides the weak synchronization feature, the attack robustness is another important feature of \ours. Existing work models the acoustic signal propagation to compensate for the propagation loss over the air~\cite{schonherr2020imperio}. But the heavy computation prevents them from being adopted in real time attack. 
Also, the quality of perturbation relies on the speaker's amplifier, and the additional distortion on such small perturbation is hard to model. Inspired by the prior work~\cite{li2020advpulse} who sets a frequency filter to guarantee the generated perturbation is ranging from 50-8,000 Hz. To guarantee the effectiveness of \ours over the air, we follow their approach on configuring a frequency filter to mitigate the uneven frequency response caused by the hardware imperfection of the speaker, thereby enhancing the attack robustness.

\section{Evaluation}

\label{sec-evaluation}
In this section, we first introduce our benchmark experimental setting to generate AEs and perturbations. Then, we evaluate PhantomSound thoroughly to validate its feasibility and robustness. Moreover, we measure the impacts of different parameters in tuning a successful attack. Our attack is successfully launched on four different ASR service APIs, and the five popular commercial IVC devices. We further conduct an user case study in section~\ref{sec-usercase}. This section describes the results in detail.

\subsection{Target Model Selection}

Since we are developing a general approach to generate perturbations to attack closed-source ASR systems and commercial devices, we will examine the effectiveness of AEs and perturbations on the most popular IVC devices available on the market. Specifically, we select Google Home (G-H), Google Assistant (G-A), Amazon Echo, Microsoft Cortana, and IBM WAA\footnote{WAA represents ``Wav-Air-API". As IBM does not own a commercial voice assistant device, we record and replay our AEs over-the-air, and transcribe them with IBM Watson API. This process, named as WAA, simulates an IVC device that is integrated with an IBM Watson API~\cite{chen2020devil}.}
as target IVC devices. Moreover, we target their respective ASR APIs, namely, Google Cloud Speech-to-Text API, Microsoft Azure API, Amazon Transcribe API, and IBM Watson API. As for Apple Siri, since there is no online speech-to-text API service available from Apple, we cannot perform \ours due to the lack of querying feedback from its ASR system. For all the target systems, we only receive the hard label of the querying input from their APIs. 

\subsection{Metrics}

We use the following metrics to quantify the effectiveness of our attack: \emph{(1) Success Rate:} this metric represents the ratio of successful attacks and the total attempts. For an untargeted attack, as long as the AEs and the perturbations alter the prediction of the original input, we count it as successful. For a targeted attack, we report success only when the prediction matches the targeted class.
\emph{(2) Average queries per command:} we use the number of queries to imply the cost and speed of AE generation. 
Specifically, we measure how many queries it needs to craft a  perturbation. This metric is calculated by the total number of queries over the number of crafted AEs/perturbations.
\emph{(3) L2 Distortion:} the L2 distortion $||\delta||^2$ indicates the size of perturbations. Prior to the launch of a physical attack, we can measure the distortion value by summarizing the squared amplitude of the generated perturbations. Note that the perturbation  $\delta \in [0,1]^l$ and the initial phoneme-level distortion ranges from 50 to 1,600 depending on different phonemes, which will be optimized after the perturbation fine-tuning as shown in Section~\ref{subsec:attack}. \emph{(4) False Accept Rate:} the false accept rate is measuring the probability of that the attacks can be false accepted by the liveness detection methods. We use this metric to evaluate the ability of our attacks to bypass the existing defense methods (e.g. liveness detection) compares to the existing attacks. The higher false accept rate we achieve, indicating the more dangerous of attack is, to bypass the existing liveness detection methods.

\subsection{Dataset}
The dataset we choose as original input is speech commands v0.02~\cite{warden2018speech} released by Google Brain. This dataset is designed to validate the keyword detection capability of DNN models. It contains 105,829 utterances of 35 common one-word commands (e.g., ``yes", ``learn", ``stop"), which is recorded from 2,618 volunteers. To validate the effectiveness of PhantomSound on a longer command, we record 10 longer commands (partially listed in Table~\ref{tab:targeted}) from a volunteer. 

For the phoneme dataset, we expect to obtain all 44 pure English phonemes with flexible duration. 
Existing speech datasets (e.g., Arabic Speech Corpus~\cite{arabic}, TIMIT~\cite{timit}) include the annotations of phonemes, but it requires extra efforts to extract individual phonemes with different duration from the speech audio. Besides, PCVC dataset~\cite{pcvc} only involves 12 volunteers, and scikit phoneme dataset~\cite{scikit} only contains 5 vowels. To construct a phoneme dataset with a diverse set of speakers, 
we use 200 different audios from 200 speakers in speech commands v0.02, remove the silence in the recordings, and randomly cut audio clips with a duration between 50ms to 300ms. This phoneme processing step follows that of the scikit phoneme dataset~\cite{scikit}, which results in 453 audio clips in total.

\begin{table}[h]
\caption{Dataset description (``unique cmds" refers to the number of unique target commands, and ``total audios" refers to the total number of (adversarial) audios that lead to the target commands).}
    \label{tab:dataset}
    \centering
\scalebox{0.8}{
\begin{tabular}{l|l|l|l|l}
\hline
      & Phone. & Cmd. & Untargeted & Targeted \\
      \hline
Unique cmds & -       & 45    & 1785     & 64     \\
\hline
Total audios & 453     & 300   & 6,219      & 216     \\
\hline
\end{tabular}
}
\end{table}

Table~\ref{tab:dataset} records the number of involved data including phonemes, commands, untargeted perturbations, and targeted perturbations. We use 35 one-word commands from the speech commands v0.02 dataset, along with 10 self-recorded long commands to build a command dataset with 45 different commands, including 300 audios in total. Then, we apply the proposed algorithm to randomly generate AEs and perturbations for an untargeted attack, resulting in 1785 different commands and 6,219 adversarial audios on 4 different commercial APIs. For the targeted attack, we attempt the perturbation of keywords, 
and generate 64 target commands with 216 adversarial audios.

\subsection{Experiment Setting}
\label{sec:setting}
We conduct the experiments on a desktop with Intel i7-7700k CPUs, 32GB RAM, and 64-bit Ubuntu 18.04 LTS operating system. The experiments are performed at three locations with different noise floors. We use three loudspeakers, including LG monitor built-in speaker (at the apartment), an SADA D6 home small speaker (at the lab), and an Samsung S9 phone (at outdoor), to transmit AEs (i.e., AE attack) and perturbations (i.e., perturbation attack) to the victim devices. Figure~\ref{fig:picture} demonstrates the attack scenario: the victim speaks commands into a smartphone or Google Home mini, while the attacker plays the perturbation through a speaker.

\subsection{Attack Performance}\label{subsec:attack}
We first evaluate the functionality of AE generation in PhantomSound. The purpose of this evaluation is 1) to demonstrate that the perturbation amplitude is negligible compared with the input, and 2) to prove the 
query efficiency of our phoneme-level searching algorithm.  
Then, we conduct the physical attack and validate the robustness of our attack over the air.

\noindent\textbf{Attack Over-the-line:} We first evaluate the attack by targeting the ASR APIs. The adversarial audios are directly supplied to the online APIs. 
We randomly select 20 adversarial audios from every command, and then perform the untargeted attack by searching for 100 epochs ($K=100$ in Algorithm~\ref{al:init}). Then, the generated perturbations are optimized to suppress their power. In the end, we obtain 148 AEs and perturbations from $\sim$44k queries ($Q=30$ in Eq.~(\ref{eq:sign-opt})), i.e., 301 queries per AE on average. 

\vspace{-10pt}
\begin{figure}[h]
\centering     
\subfigure[One-word commands  ]{\label{fig:short}\includegraphics[width=40mm]{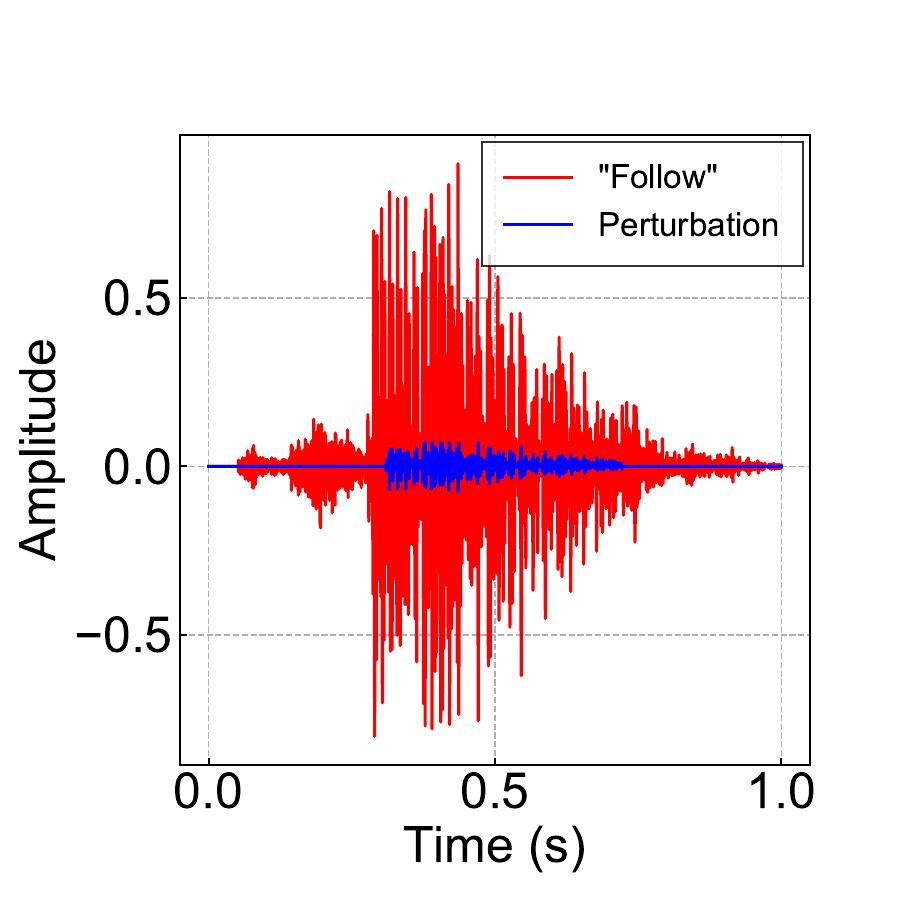}}
\subfigure[Command phrases]{\label{fig:long}\includegraphics[width=40mm]{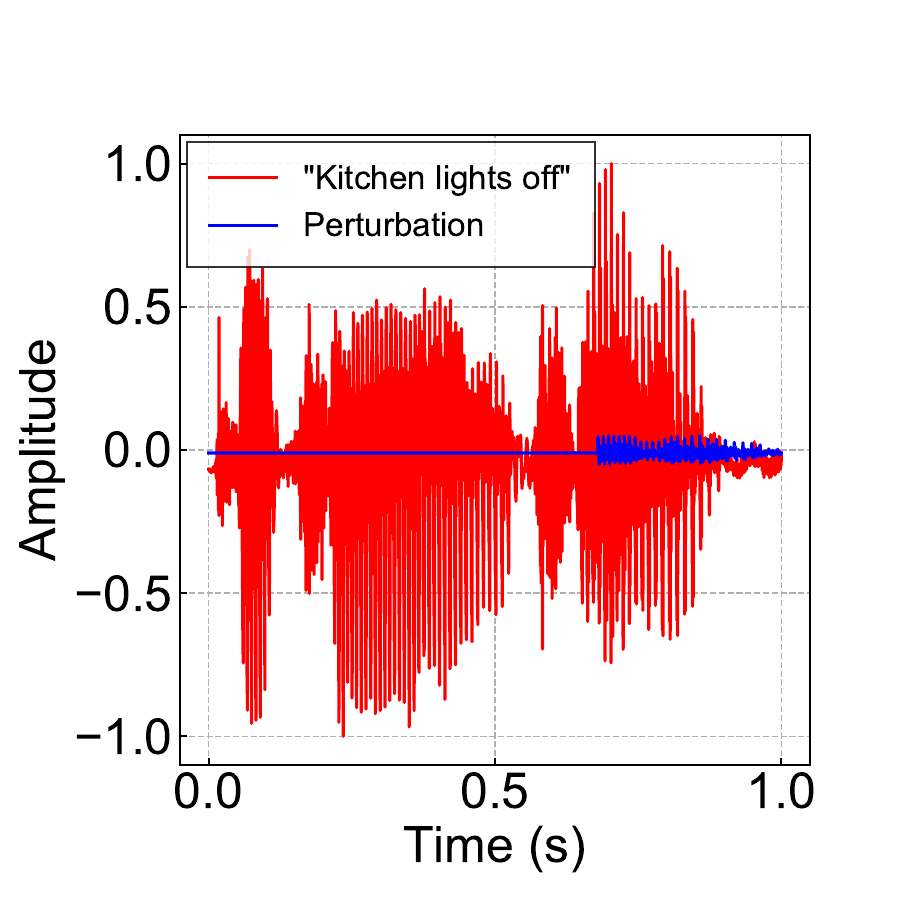}}
\vspace{-15pt}
\caption{Comparison of input and perturbation amplitudes.}
\label{fig:amp_comp}
\end{figure}

To evaluate the perturbation amplitude, we randomly pick two examples from the generated perturbation as shown in Fig.~\ref{fig:amp_comp}. 
We can see that the crafted perturbations have a negligible power profile compared with the input regardless of the length of commands. Moreover, the duration of perturbation is shorter than the input, which makes it possible to conceal the presence of perturbation. 
Table~\ref{tab:untarget_command} summarizes the results of the untargeted attacks toward 4 types of commercial APIs. We observe that every command can be altered into at least two false commands. While some of the false predictions are harmless, the attack can almost certainly 
\emph{invalidate} the victim's command. Moreover, in certain cases, some perturbations can lead to a contrary response from voice assistants (e.g., ``right" to ``wrong" in Amazon Transcribe API, ``right" to ``no" in Microsoft Azure API). Considering the number of queries for generating one perturbation, the Google Cloud Speech-to-Text is reported to be the most resilient API under our attack, as it requires the most queries. 

\vspace{-5pt}
\begin{table}[h]
\caption{Untargeted attack results.}
    \label{tab:untarget_command}
    \centering
\scalebox{0.8}{
\begin{tabular}{c|l|l|l|l}
\hline
\hline
\textbf{Cmds.}             & \begin{tabular}[c]{@{}l@{}}Google \\ Cloud\end{tabular} & \begin{tabular}[c]{@{}l@{}}MS \\ Azure\end{tabular} & \begin{tabular}[c]{@{}l@{}}AMZ\\ Trans.\end{tabular} & \begin{tabular}[c]{@{}l@{}}IBM\\ Watson\end{tabular} \\
\hline
\multirow{3}{*}{"down"} & "damn"  &  "town"  &   "done"  &   "Downer"   \\
                     & "done"   &  "one"  &    "dine" &    "Done"\\
                     & "does"  & "south"    &  "drive" &   "Drone"\\
                     \hline
\multirow{3}{*}{"follow"}& "fallout" &   "fallout"    &  “no”  &    "fallen"\\
                     & "farm"     &     "fall over"    &   "for sure"       &     "fall over"\\
                     & "four"      &      "learn"   &  "phone"              &   "fall" \\
                     \hline
\multirow{3}{*}{"forward"} & "forewarn" &   "work"   &    "what"    &     "for"\\
                     & "for eyes"    &  "for"    &  "work"   &       "four"\\
                     & "for work"    &  "ford"       &   & "for all"\\
                     \hline
\multirow{3}{*}{"yes"} & "yeah"      &  "file"    &     "yeah"   &     "yeah"\\
                     & "yeah!"    & "4"     &   "yes.."  &      "yet" \\
                     & "yet"    &   "On"      &      "right"   &  "hi"\\     
                     \hline
\multirow{3}{*}{"right"} & "Rite Aid"      &  "no"    &  "write"      &    "run" \\
                     & "write"    &   "go"   &    "run" &       "ray"\\
                     & "read"    &    "trade"     &      "wrong"   &    "left" \\  
                     \hline
                     \hline
\textbf{Queries} & 345 & 251  &  215 & 314 \\
                     \hline
                     \hline
\end{tabular}}
\end{table}

To further comprehend the query effectiveness, we conduct an additional experiment to validate the sensitivity of different APIs in terms of request rejection rates. The result shows that Google API is most sensitive as it refuses to respond to an unclear input, while the Amazon transcribe always responds to any inputs. 
Table~\ref{tab:targeted} records the targeted attack results towards a longer input. 
\begin{table}[h]
\caption{Targeted attack results}
\label{tab:targeted}
\centering
\scalebox{0.7}{
\begin{tabular}{l|c|llll}
\hline
\hline
\multicolumn{2}{c}{Command}     & \multicolumn{4}{c}{Query}      \\
\hline
\multicolumn{1}{c}{Input}          & Target     
& \begin{tabular}[c]{@{}l@{}}Google \\ Cloud\end{tabular} & \begin{tabular}[c]{@{}l@{}}MS \\ Azure\end{tabular} & \begin{tabular}[c]{@{}l@{}}AMZ\\ Trans.\end{tabular} & \begin{tabular}[c]{@{}l@{}}IBM\\ Wat.\end{tabular} \\
\hline
"turn \textbf{right}"           & "turn \textbf{left}"       & 1,895     & 1,128    & 1,421    & 1,487   \\
\hline
\begin{tabular}[c]{@{}l@{}}"kitchen \\ lights \textbf{off}"\end{tabular} & \begin{tabular}[c]{@{}l@{}}"kitchen \\ lights \textbf{on}"\end{tabular}    & 1,754       & 857     & 933    & 1,377  \\
\hline
"call \textbf{mom}"     & "call \textbf{911}"     & -    & 1,421   & 1,125   & - \\
\hline
\begin{tabular}[c]{@{}l@{}}"\textbf{read} \\ my \\ message"\end{tabular} & \begin{tabular}[c]{@{}l@{}}"\textbf{delete} \\ my \\ message"\end{tabular} & 2,342  & 1,520  & 1,436   & 1,781         \\\hline
\multicolumn{2}{c}{\begin{tabular}[c]{@{}l@{}}\textbf{Average}\\ \textbf{Queries}\end{tabular}}    & 1,997      & 1,232     & 1,229    & 1,548      \\        \hline
\hline
\end{tabular}
}
\end{table}
The results show that our phoneme-level searching method is capable of finding the specific perturbation that could mislead the APIs to return a target result. 
Note that the average query amount increases dramatically in the targeted attack case, which is anticipated because the target need to be achieved by multiple round perturbation searching (line 7-10 in Algorithm~\ref{al:init}). It is also noteworthy that our targeted attack cannot guarantee finding a successful perturbation under any arbitrary inputs (e.g., Google Cloud fails to craft AEs for ``call 911").

\begin{table}[h]
\caption{Comparison for Untargeted Attacks}
\centering
\scalebox{0.8}{
\begin{tabular}{l|l|l|l|l}
\hline
Models$\downarrow$ &  Ours & \begin{tabular}[c]{@{}l@{}}white\\ box\cite{carlini2018audio}\end{tabular} & \begin{tabular}[c]{@{}l@{}}score\\ based\cite{chen2017zoo}\end{tabular} & \begin{tabular}[c]{@{}l@{}}brute \\ force\cite{brendel2017decision}\end{tabular} \\

\hline
DS 1~\cite{hannun2014deep} & \textbf{185}  & \textbf{90}                                                   & 206                                                      & $\infty$                                                            \\
\hline
DS 2~\cite{amodei2016deep} & \textbf{226}  & \textbf{75}                                                    & 197                                                      &          $\infty$                                                             \\
\hline
\end{tabular}
}
\label{tab:known_query}
\end{table}

\noindent\textbf{Query Efficiency Comparison on Known models:} To validate the benefits of introducing phonemes to guide the optimization direction, we implement 3 different attacks on two known models. 
By attacking two ASR models (DeepSpeech 1/DS 1~\cite{hannun2014deep} and DeepSpeech 2/DS 2~\cite{amodei2016deep}) with different prior knowledge and method, we find that \ours achieves comparable query efficiency with the grey box setting, with 100\% attack success rate. The result is summarized in Table~\ref{tab:known_query}. Given the same 10 benign commands, we use the 4 attacks to generate untargeted AEs with the same $L_2$ distortion. We record the average number of queries for different prior knowledge of the victim model. Compares to the white box attack, which can fine-tune in <90 queries, \ours requests ~200 queries to craft an AE, which is close to the queries of a score-based attack. This result indicates that our strategy such as 1) using phoneme to initialize perturbation 2) Query-efficient fine-tuning is working well, and performing similar results with less information (e.g., confidence score). It is noteworthy that the brute force decision boundary search method doesn't work for attacking the ASR model. Because this method initializes a random noise and retrieves model gradients by altering the noise. However, this noise can never be fine-tuned while the victim model produces an empty label to it, resulting in an infinite number of queries.

\begin{table}[h]
\caption{Comparison for Targeted Attacks}
\label{tab:queryeff}
\centering
\scalebox{0.8}{
\begin{tabular}{l|l|c|c}
\hline
Attacks & Knowledge  & Queries & SR \\
\hline
Carlini~\cite{carlini2018audio} & Gradient  & $\sim$1,000 & 100\% \\
\hline
Houdini~\cite{cisse2017houdini}  & Gradient  & $\sim$1,000 & 100\%\\
\hline
Devil's~\cite{chen2020devil}  & Conf. Score  & $\sim$1,500 & 100\%\\
\hline
OCCAM~\cite{zheng2021black}   & Final decision  & $\sim$30,000 & 100\%\\
\hline
Ours &\textbf{Final decision}    & \textbf{$\sim$1,500} & \textbf{68\%}\\
\hline
\end{tabular}
}
\end{table}

\noindent\textbf{Query Efficiency Comparison for Targeted Attacks:} We compare the number of required queries with four existing attacks in Table~\ref{tab:queryeff}. The white-box attacks (Wb)~\cite{carlini2017towards, cisse2017houdini} require the least amount of queries ($\sim$1,500). 
With the knowledge of confidence scores of API's decoding results, the Devil’s Whisper~\cite{chen2020devil} utilizes a surrogate model trained with around 1,500 queries to attack the APIs. In the scenario when an attacker  can  only access the final decision of the query API, \ours needs $\sim$1,500 queries (comparable with the white-box setting) to craft a targeted perturbation. Compared with a recent black-box attack OCCAM~\cite{zheng2021black}, we reduce the number of queries by 95\%. However, due to the limitation of phoneme length and diversity, we sacrifice the success rate to achieve high query efficiency.

\begin{figure}[h]
\centering     
\subfigure[Weak synchronization]{\label{fig:offset}\includegraphics[width=40mm]{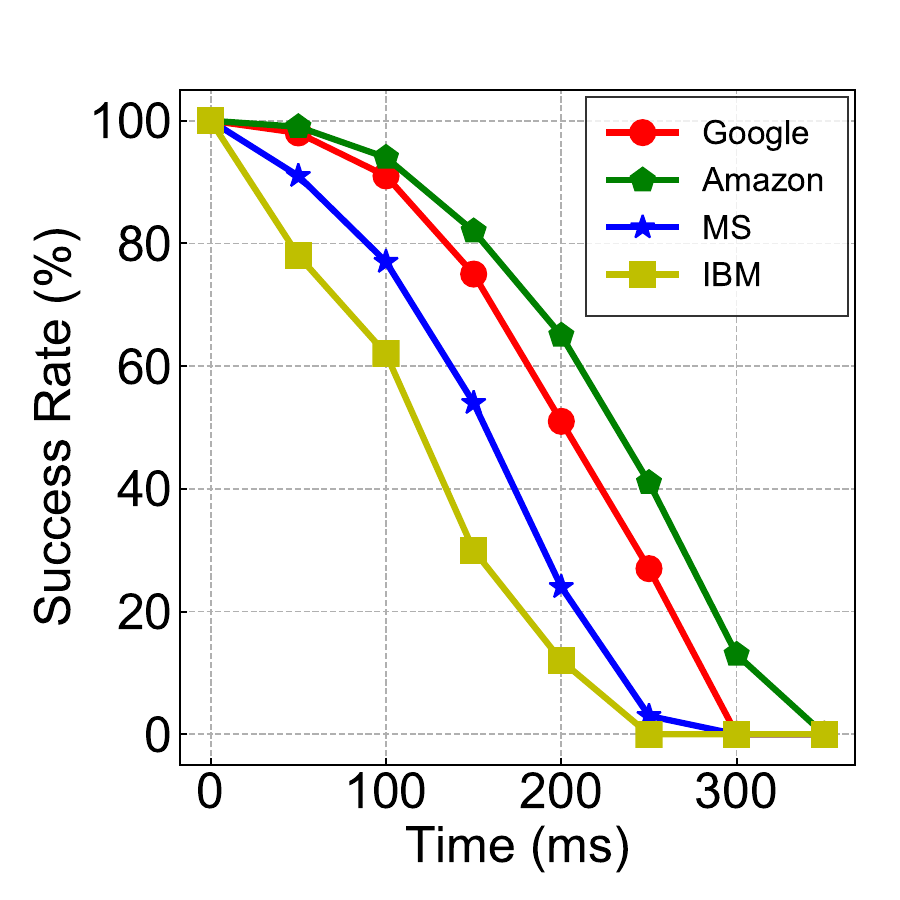}}
\subfigure[$L_2$ Distortion vs. No. of queries]{\label{fig:query}\includegraphics[width=40mm]{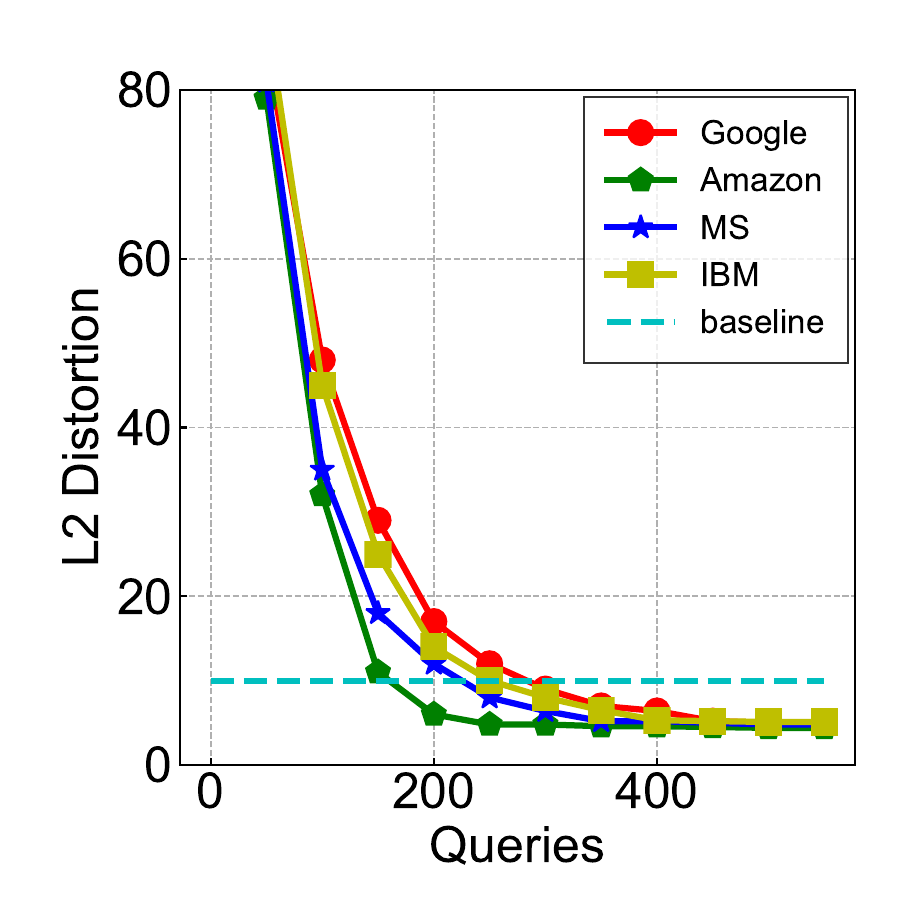}}
\vspace{-10pt}
\caption{Evaluation of AE generation process.}
\vspace{-10pt}
\end{figure}

\noindent\textbf{Weak synchronization:} 
Before evaluating the physical attacks, 
we investigate the effectiveness of the proposed weak synchronization design. 
In this experiment, we manually add mismatch delays between input $x_0$ and the generated perturbation to craft mismatched AEs. We then use the mismatched AEs to query the APIs and measure the attack success rate. Fig.~\ref{fig:offset} displays the result, from which we can see that, after using the average loss, although we expect the weak-synchronization works within $400ms$ (detailed in Section~\ref{sec:weak_sync}), this design is only partially effective, because the success rate drops steadily with the increasing mismatch time. Moreover, we show the tendency of L2 distortion w.r.t. the number of queries in Fig.~\ref{fig:query}. The baseline denotes an L2 distortion of 10, which is proven unnoticeable by two volunteers when AEs are played using an LG monitor with a medium volume. 

\begin{figure*}[t]
\centering     
\subfigure[AE attack]{\label{fig:real_ae}\includegraphics[width=0.24\linewidth]{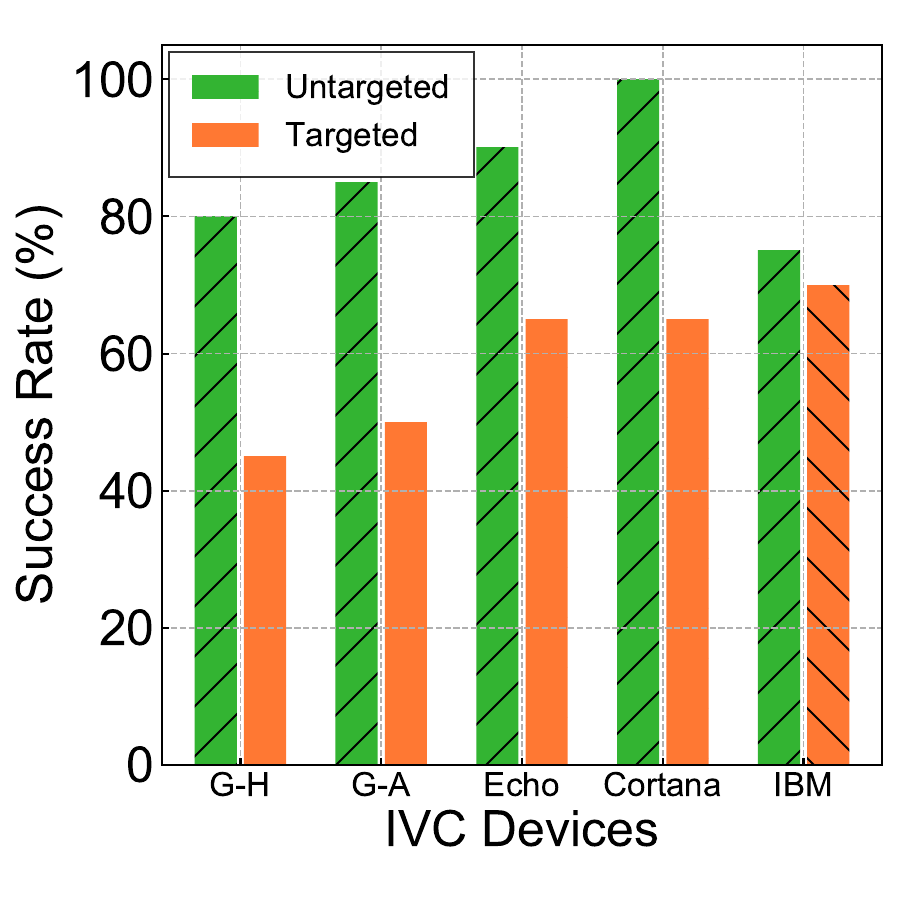}}
\subfigure[Perturbation attack]{\label{fig:real_pe}\includegraphics[width=0.24\linewidth]{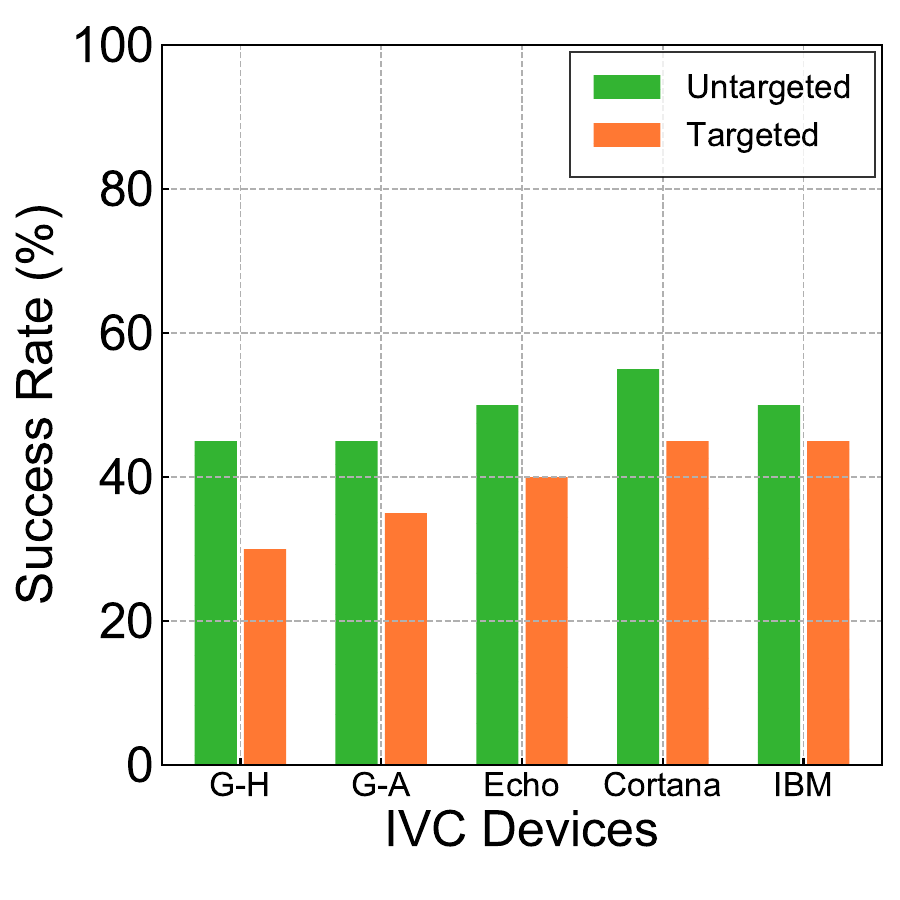}}
\subfigure[Distance]{\label{fig:dist}\includegraphics[width=0.24\linewidth]{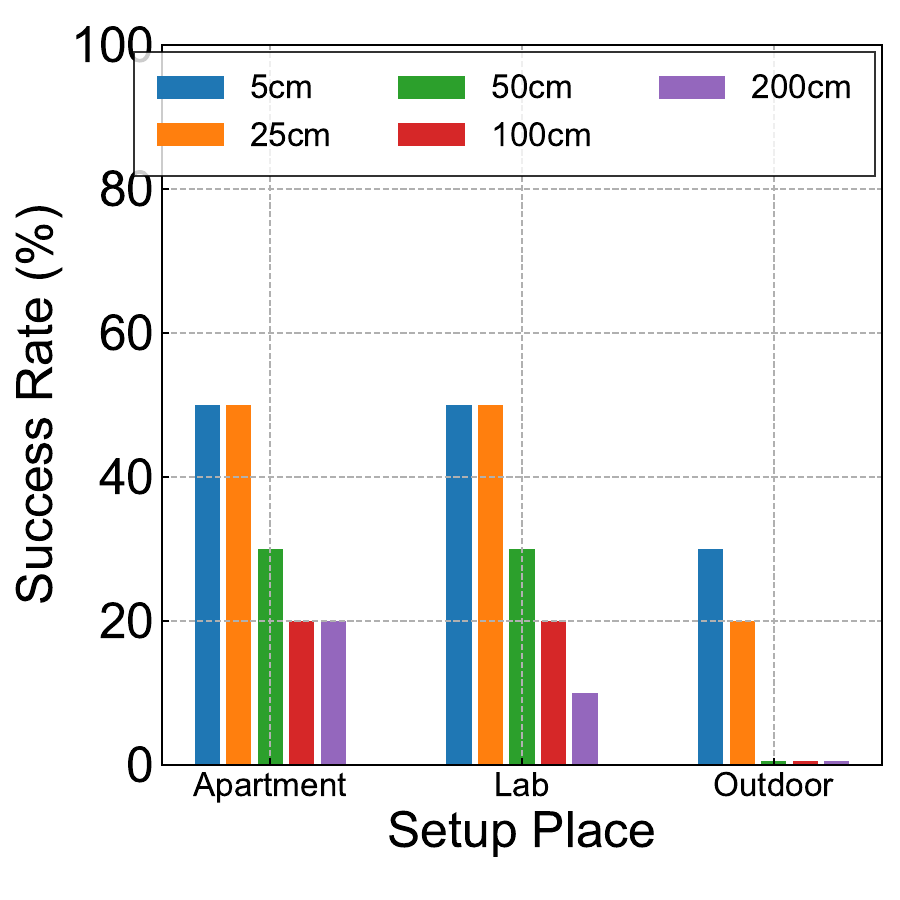}}
\subfigure[Loudness]{\label{fig:loud}\includegraphics[width=0.24\linewidth]{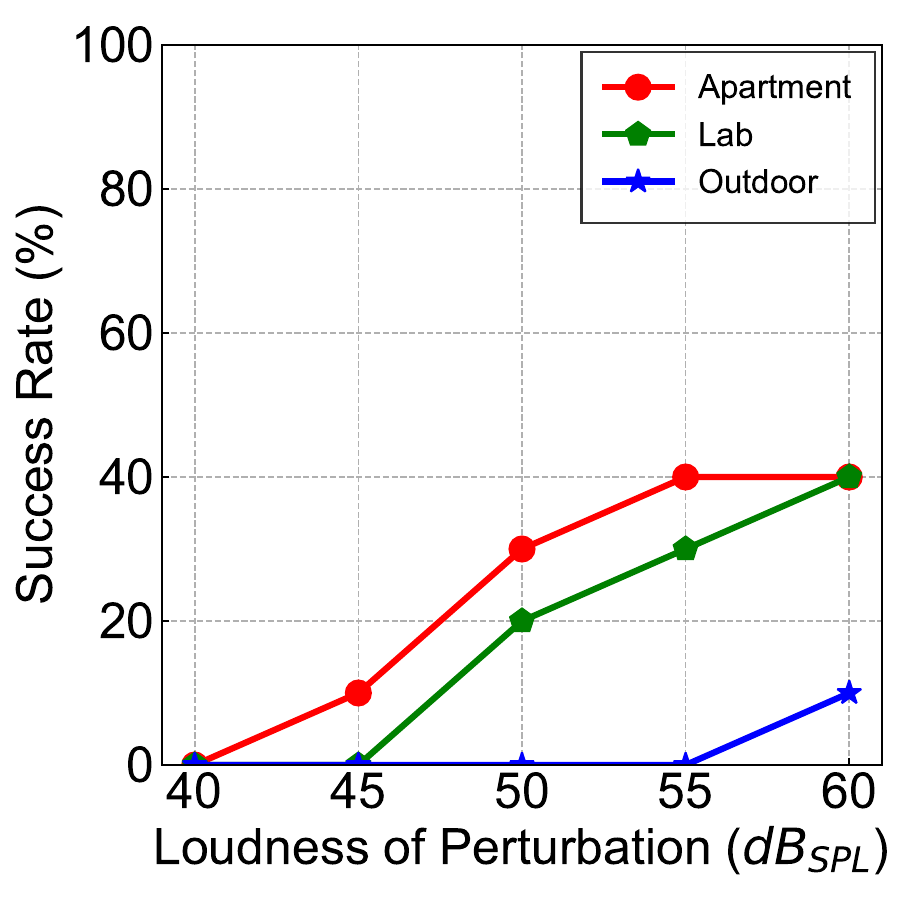}}
\vspace{-10pt}
\caption{AE generation results.}
\vspace{-10pt}
\end{figure*}
\noindent\textbf{Attack Over-the-air:} The over-the-air attack evaluation aims to prove the robustness of \ours. 

\rev{To attack commercial APIs, we play the valid AEs and perturbations (which  attack successfully in over-the-line scenarios) via a SADA D6 speaker, and record it by iPhone 12 Pro, the recordings are sent to the commercial API for evaluation. The attack distance is set to 50cm.
For each attack, we choose 5 AEs to play 5 times and get the average success rate. We report the result in Table~\ref{tab:baseline_ova}.
\begin{table}[h!]
\caption{Over-the-air attack API baseline}
\centering
\scalebox{0.8}{
\begin{tabular}{|c|c|c|c|c|c|}
\hline
                            \multicolumn{2}{|c|}{APIs} &
                            Google Cloud & 
                            MS Azure & 
                            AMZ Trans & 
                            IBM Wat.  \\
\hline
\multirow{2}{*}{Targeted} & AE & 76\% & 80\% & 80\% & 84\% \\
\cline{2-6}
                            & Pert. & 68\% & 72\% & 72\% & 76\% \\
\hline
\multirow{2}{*}{Untargeted} & AE & 100\% & 100\% & 100\% & 100\% \\
\cline{2-6}
                            & Pert. & 72\%  & 80\% & 80\% & 92\% \\
\hline
\end{tabular}
}
\label{tab:baseline_ova}
\end{table}
From our observations, it is apparent that in the context of a targeted attack, our method attains approximately a 
$\sim 80\%$ success rate in attacking over-the-air commercial ASR APIs by directly playing the audio adversarial example (AE). When the attack is synchronized with the victim's speech, the perturbation attack exhibits around a $\sim72\%$ success rate. On the other hand, when it comes to untargeted attacks, our adversarial examples (AE) and perturbation methods achieve impressively high success rates. They misdirect the victim's input with a $100\%$ and approximately $81\%$ success rate, respectively.
Next, we follow the same setting to attack commercial IVC devices.

The result in Fig.~\ref{fig:real_ae} uncovers the success rate of playing AEs directly. Among all the tested IVC devices, Microsoft Cortana is most vulnerable against the AE attack, while the Google series products (e.g., Google Home, Google Assistant) show the most resilience 
against the targeted AE attack. Overall, the success rate of an untargeted attack is higher than that of a targeted one, i.e., the former reaches $\sim$80\% success rate and the latter stays around $\sim$50\%. 
With the perturbation attack, Fig.~\ref{fig:real_pe} reveals a relatively low success rate. Similarly, compared to the targeted perturbation attack, the untargeted attack has a higher success probability, achieving around 45\% success rate on average. Nevertheless, the success rate can be further improved via multiple repeated attempts. 
We also summarize the success rate compared to prior black-box attacks in Table~\ref{tab:time_use}. 

\begin{table}[h] 
\caption{Comparison with other real-world attacks}
\label{tab:time_use}
\centering
\scalebox{0.7}{
\begin{tabular}{l|l|l|l|l|l|l|l|l}
\hline
\begin{tabular}[c]{@{}l@{}}Target\end{tabular} 
& \begin{tabular}[c]{@{}l@{}}Google\\ Cloud\end{tabular} & \begin{tabular}[c]{@{}l@{}}MS\\ Azure\end{tabular} & \begin{tabular}[c]{@{}l@{}}AMZ\\  Trans.\end{tabular} & \begin{tabular}[c]{@{}l@{}}IBM\\ Wat.\end{tabular} &
\begin{tabular}[c]{@{}l@{}}Google\\ Home\end{tabular} & 
\begin{tabular}[c]{@{}l@{}}Google\\ Assit.\end{tabular} & \begin{tabular}[c]{@{}l@{}}MS\\ Cortana\end{tabular} & \begin{tabular}[c]{@{}l@{}}AMZ\\ Echo\end{tabular} \\ 
\hline
Devil's~\cite{chen2020devil}&10/10&10/10&4/10&10/10&9/10&10/10&10/10&10/10\\
\hline
Danger~\cite{zhang2019dangerous}&-&-&-&-&15/100&-&-&69/100\\
\hline
Ours&19/25& 20/25 & 20/25 & 21/25 & 11/25 & 12/25 & 16/25 &16/25\\
\hline
\end{tabular}
}
\end{table}

Upon comparison with the Devil's attack~\cite{chen2020devil}, it is evident that our attack method yields a marginally lower success rate against the APIs, with the exception of the Amazon Transcribe API. Considering the IVC devices, the Devil's attack tends to be more effective at similar SNR levels. For the Danger attack~\cite{zhang2019dangerous}, we have displayed their success rate derived from their "voice squatting" attack, where the victim's command is misinterpreted to initiate the attack skill. A comparison reveals that our attack technique yields comparable success rates when targeting Amazon Echo, and even demonstrates superior performance when used to attack Google Home.
}


\begin{table}[h] 
\caption{Latency for perturbation generation}

\label{tab:time_use}
\centering
\scalebox{0.8}{
\begin{tabular}{l|l|l|l|l}
\hline
\begin{tabular}[c]{@{}l@{}}Time\\ Consumption\end{tabular} & \begin{tabular}[c]{@{}l@{}}Google\\ Cloud\end{tabular} & \begin{tabular}[c]{@{}l@{}}MS\\ Azure\end{tabular} & \begin{tabular}[c]{@{}l@{}}AMZ\\  Trans.\end{tabular} & \begin{tabular}[c]{@{}l@{}}IBM\\ Wat.\end{tabular} \\ \hline
Latency (s)                                                & 0.29                                                   & 0.58                                               & 26.31                                                 & 1.35                                               \\ \hline
Untargeted (min)                                           & 1.67                                                   & 2.43                                               & 94.3                                                  & 7.1                                                \\ \hline
Targeted (min)                                             & 9.65                                                   & 11.9                                               & 539                                                   & 34.8                                               \\ \hline
\end{tabular}
}
\end{table}

\noindent\textbf{Time Cost:} Different from the prior works that require a substantial amount of time to craft AEs offline, \ours enables much faster AE generation. Such a fast generation feature is essential in practice, when the attackers only have a limited time budget to instantiate the attack. 

In the experiment, we record the latency for querying 4 different commercial APIs to get the results.  
The results are presented in the first row of Table~\ref{tab:time_use}, which show that 3/4 of APIs could return a result in seconds, except Amazon Transcribe API. The Amazon API has to interact with Amazon Web Service and Storage bucket, which spends a longer period for the results to return. 

We then compute the total time needed for perturbation generation, by multiplying latency with the number of queries (shown in Table~\ref{tab:untarget_command},~\ref{tab:targeted}). Our result shows that
PhantomSound can generate a perturbation for both the targeted and untargeted attacks in minutes with the exception of Amazon API, while the targeted one takes longer. 
Note that we take the $L_2$ distortion into consideration during the time cost computation, however, if the attacker ignores the impact of the perturbation loudness and uses the intermediate perturbation, the generation time can be further reduced. 

\begin{figure}[h]
    \centering
    \includegraphics[width=0.7\linewidth]{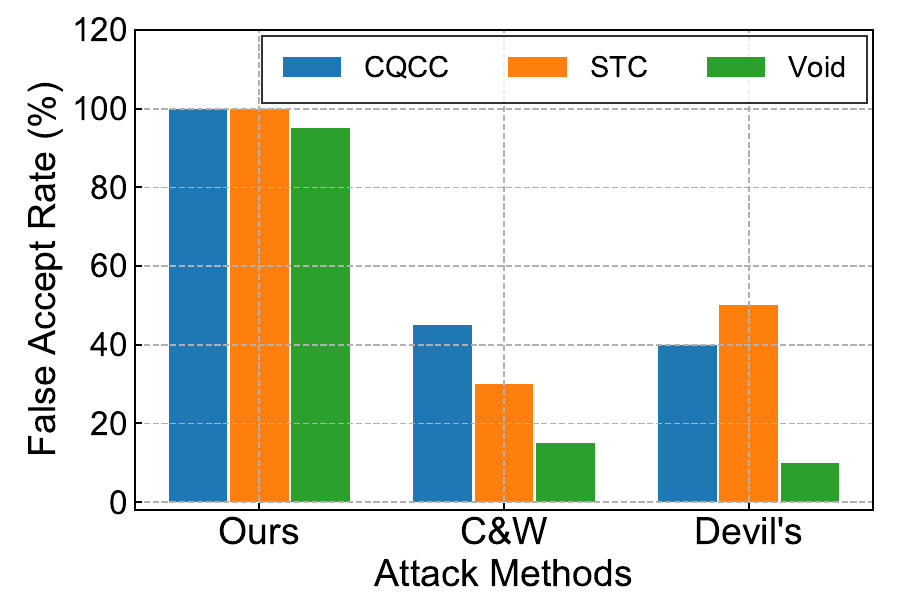}
    \vspace{-10pt}
    \caption{Attacks vs. liveness detection defenses}
    \label{fig:bypass_liveness}
    \vspace{-15pt}
\end{figure}

\subsection{Ability to Bypass Liveness detection}
Compares to the existing physical adversarial attacks~\cite{yuan2018commandersong, chen2020devil, zheng2021black}, \ours relies on the benign commands spoken by the user. Although this attack setting requires extra effort to synchronize the perturbation and the user's benign speech, it brings potential benefits to bypassing the defense mechanism. For example, recent works~\cite{ahmedvoid, mengyour, li2021robust, kinnunen2017asvspoof, lavrentyeva2017audio, guo2022supervoice} proposed liveness detection approaches can differentiate the source of sound (human or machine) with high accuracy. Therefore, the conventional adversarial attacks that are launched solely by loudspeaker~\cite{yuan2018commandersong, chen2020devil, zheng2021black} have a higher probability to be defended by those liveness detection methods. In contrast, our attack is designed to launch with the user's speech, leading to a more dangerous threat to the liveness detection defenses. To validate the performance of \ours over different defense mechanisms, we reproduce three liveness detection algorithms, CQCC~\cite{kinnunen2017asvspoof}, STC~\cite{lavrentyeva2017audio}, and Void~\cite{ahmedvoid}; For comparison, we implement  C\&W attack~\cite{carlini2018audio} and Devil's~\cite{chen2020devil}to attack with liveness detection algorithms.
The detailed liveness detection methods can be found in Appendix~\ref{sec-live}.
To conduct this experiment, we follow the settings described as follows:

\noindent
\textbf{Ours:} We play our perturbation when the user gives the command, and record it with a smartphone. Then, we run three liveness detection algorithms to detect the sound source.

\noindent
\textbf{C\&W~\cite{carlini2018audio}:} We play the AEs that are generated by this attack, and then record with the same smartphone and run liveness detection algorithms to defend it.

\noindent
\textbf{Devil's~\cite{chen2020devil}:} We play the AEs provided from the paper's demonstration website, and then record it with the same smartphone, followed by the same liveness detection procedure.

For our attack and the C\&W attack, we use 20 different perturbations/AEs to attack the liveness detection model; As for the Devil's attack, since we can only collect 10 AEs from the demonstration website, we use 10 AEs to attack the liveness detection model. We present our result in Fig.~\ref{fig:bypass_liveness}. It is evident to show that our attack can bypass the three liveness detection models, resulting 95\% to 100\% false accept rate. In contrast, the other two attacks have a very low chance to counter the Void~\cite{ahmedvoid} detection with less than 15\% FAR. Even for conventional liveness detection methods (e.g., CQCC and STC), the existing attacks that use complete AEs also have a low probability (~40\%) to attack successfully.

\subsection{Impact of Practical Factors}\label{subsec:impact}
To investigate the critical factors that may affect the success rate of \ours, we evaluate the perturbation attack under different environments (e.g., apartment, lab, outdoor). The ambient noise level for the aforementioned places are 39.8 $dB_{SPL}$ (apartment), 41.2 $dB_{SPL}$ (lab) and 58 $dB_{SPL}$ (outdoor), respectively. 

In this experiment, we play a crafted perturbation of ``turn right" 10 times, attempting to transform the prediction into ``turn left", and the volume of perturbation is 60 $db_{SPL}$. We then record the success rate under different circumstances. Fig.~\ref{fig:dist} demonstrates the impact of attack distances, i.e., the closer the adversary is, the higher success rate he/she achieves, which is unsurprising given that our attack relies on the successful delivery of the perturbation. The relatively short attack distance is in fact a common limitation reported by the existing work~\cite{chen2020devil, yuan2018commandersong, li2020advpulse}. 
However, the attacker can further extend the attack distance by increasing the speaker's volume (though it could make the perturbation more noticeable) or utilizing a speaker array~\cite{roy2018inaudible}. 
Next, we provide the results on how the loudness factor could affect the attack performance in Fig.~\ref{fig:loud}. We can see that the success rate improves with the increasing perturbation loudness. This result also coincides with the prior work~\cite{li2020advpulse}. In an outdoor environment, it is suggested that the adversary enhance the attack robustness by 
amplifying the perturbations. Due to the higher noise level outdoors, the phoneme-like perturbation can still be hard to perceive.

\begin{figure}[h]
\centering     
\subfigure[Attack Angle iPhone 12Pro]{\label{fig:picture}\includegraphics[width=0.4\linewidth]{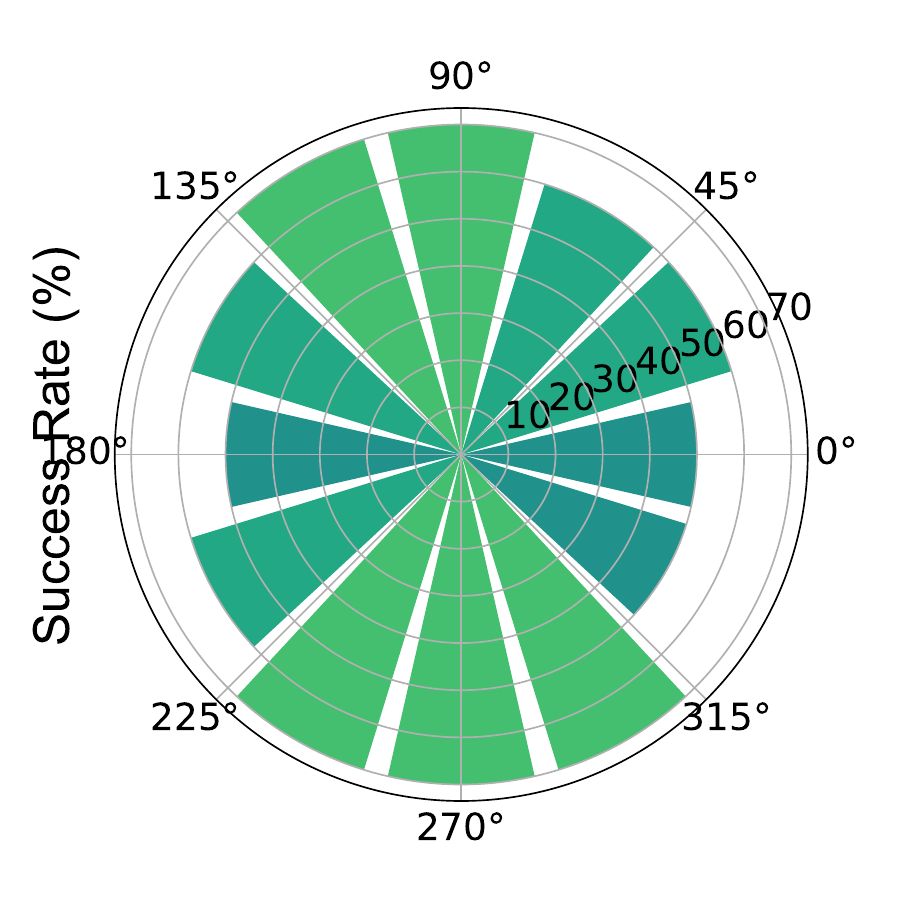}}
\subfigure[Attack Angle Mi 8 Lite]{\label{fig:perception}\includegraphics[width=0.4\linewidth]{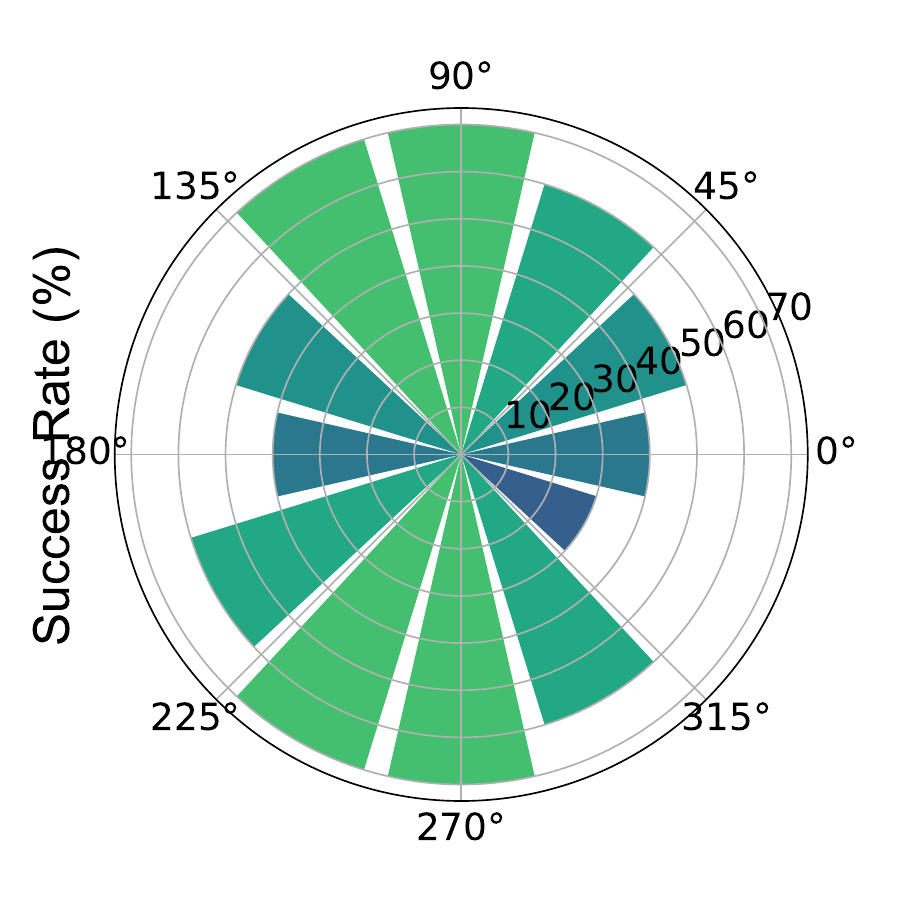}}
\vspace{-10pt}
\caption{Attack with different angles}
\vspace{-10pt}
\label{fig:angle}
\end{figure}

\noindent
\textbf{Impact of Attack Angles:} Besides the attack environments and the distance, the attack angle can also alter the attack performance. We evaluate our attack by playing AEs to two smartphones in 12 different directions (from 0 degrees to 360 degrees, with 30-degree intervals). This experiment is conducted in Lab environment and attacks the google assistant on the smartphone. We play 10 AEs in every direction with 60$dB_{SPL}$, and record the success rate of the untargeted attack. We report our result in Fig.~\ref{fig:angle}. We find that our attack has the best performance when the adversary is facing or back to the smartphone. While attacking through the side direction (e.g., 0 degrees when the adversary is parallel to the victim), the success rate is impaired. We observe the same trend on two smartphones. This result indicates that the microphone arrangement and its direction will lead to audio information loss. Unfortunately, the low power of our perturbation is hard to be sensed with the audio loss, therefore causing a low success rate in the side direction. 

\noindent\rev{\textbf{Impact of Different victims:}
In the attack preparation period, every perturbation is crafted based on a specific command from a specific speaker.
However, the adversary may use the crafted perturbation on the previous victim to attack the current victim. Here, we evaluate the capability of \ours to attack different speakers. First, we obtain 4 perturbations from speaker \#1 (male), which convert the benign commands "stop", "right", "yes", and "down" into 4 target commands "backward", "left", "no" and "song" respectively. Next, we randomly select 100 speakers (50 males and 50 females) who are not speaker \#1 from the speech commands v0.02 dataset, and inject the perturbations into their benign audio samples. For the targeted attack, if the benign commands are successfully interpreted as the target, we classify it as successful. For the untargeted attack, any case where the benign commands are misinterpreted is considered successful. The result is present in Fig.~\ref{fig:diff_victim}.

\vspace{-10pt}
\begin{figure}[h]
\centering     
\subfigure[Targeted attack]{\label{fig:diff_victim_t}\includegraphics[width=40mm]{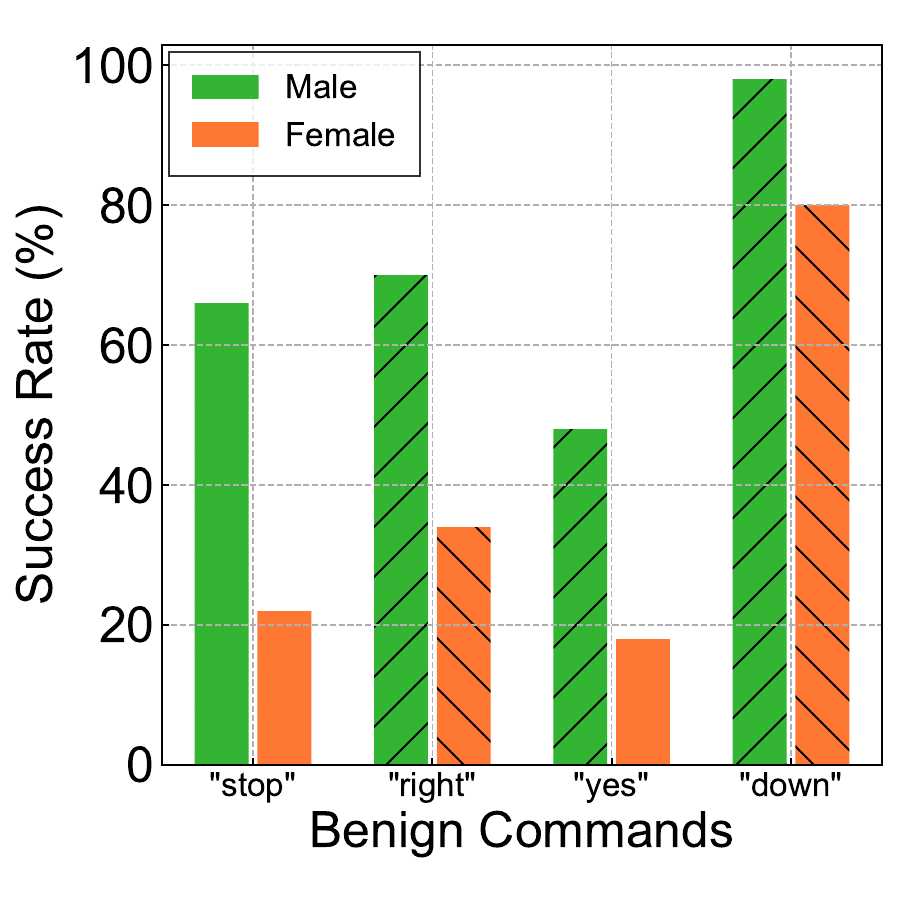}}
\subfigure[Untargeted attack]{\label{fig:diff_victim_ut}\includegraphics[width=40mm]{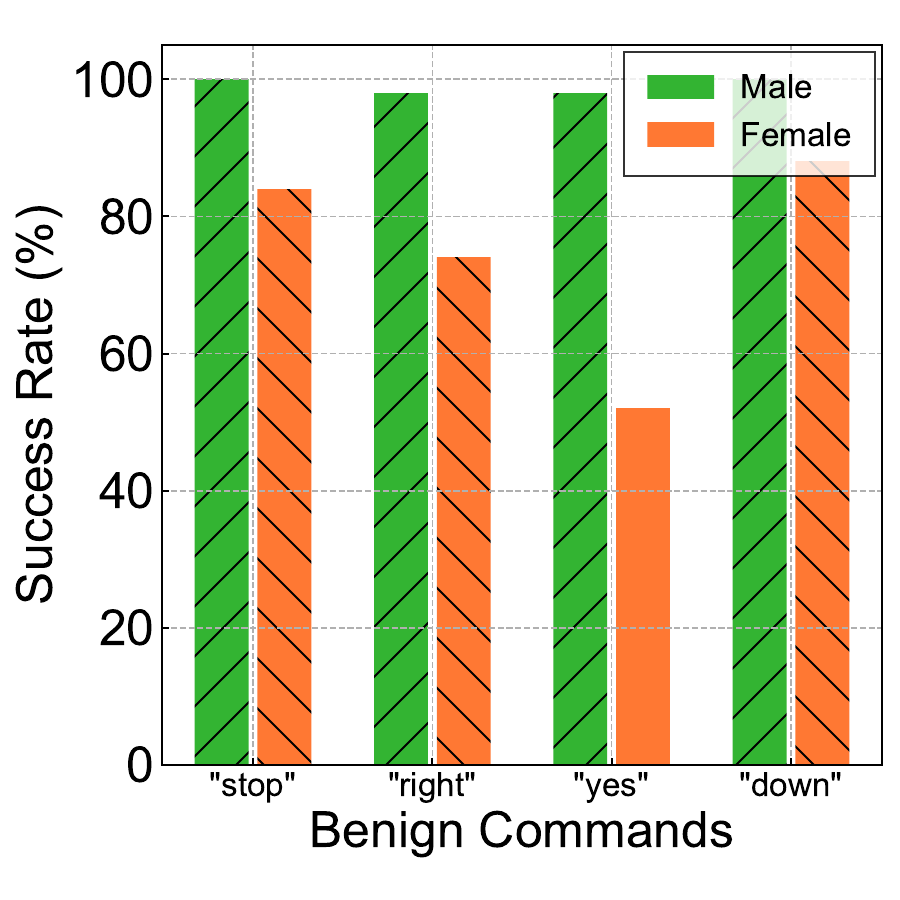}}
\vspace{-10pt}
\caption{Attack cross different victims}
\vspace{-10pt}
\label{fig:diff_victim}
\end{figure}

The result indicates that, for targeted attacks, the attack success rate is dependent on the benign samples.
The success rate exceeds $50\%$ when the target is of the same gender, but it falls below $40\%$ when targeting different genders. Regarding the untargeted attacks, the perturbations demonstrate robust transferability for attacking various speakers. The average success rate is notably high, reaching $98\%$ for males and $74\%$ for females.

}

\subsection{User Study}\label{sec-usercase}
To evaluate the stealthiness of perturbation in a real-world attack, 
we conduct an online/in-person user study to investigate the users' perception level of PhantomSound. In our study, 20 volunteers are involved, and they are requested to hear 6 crafted perturbations across 4 different distances. Two volunteers attend the in-person experiment (see Fig.~\ref{fig:picture}) and the rest of them carry out the experiment at their homes. We recruit 13 volunteers from Amazon Mechanical Turk with complete experimental instructions. The experiment setup detail can be found at Appendix~\ref{sec-ucset}.

The volunteers are asked to \emph{pretend speaking to their voice assistants} while hearing the perturbation, after which they will answer questions to depict their comprehension of the heard perturbations. The options for perception levels include: 
\emph{Listened}, \emph{Abnormal}, and \emph{Recognize}. \emph{Listened} indicates that the volunteer can hear a perturbation but regard it as a normal noise; \emph{Abnormal} implies that they hear some strange sounds; and \emph{Recognize} stipulates that they can understand the meaning of the heard sound. We report the experiment result in Fig.~\ref{fig:perception}. It shows that most of the participants can hear the perturbation within a short distance, but less than 50\% of them regard the perturbation as an abnormal sound. Such ``abnormality" feeling will gradually disappear  with the increasing attack range, which ends with 10\% in 2 meters. 
Moreover, even though all the perturbations are ``meaningless" phonemes, some participants claim to understand their meanings (though the understanding is incorrect). To summarize, \ours can be noticed by victims, but would not vastly raise their attentions. Notably, the victims are generally unaware of the meaning of perturbations.   

\begin{figure}[h]
\centering     
\subfigure[Experiment setup]{\label{fig:picture}\includegraphics[width=0.4\linewidth]{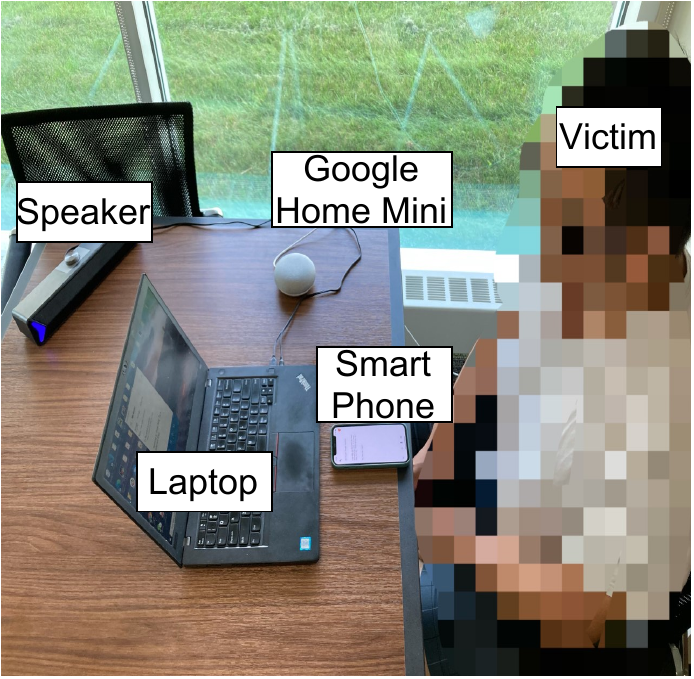}}
\subfigure[Users' perception level of \ours]{\label{fig:perception}\includegraphics[width=0.4\linewidth]{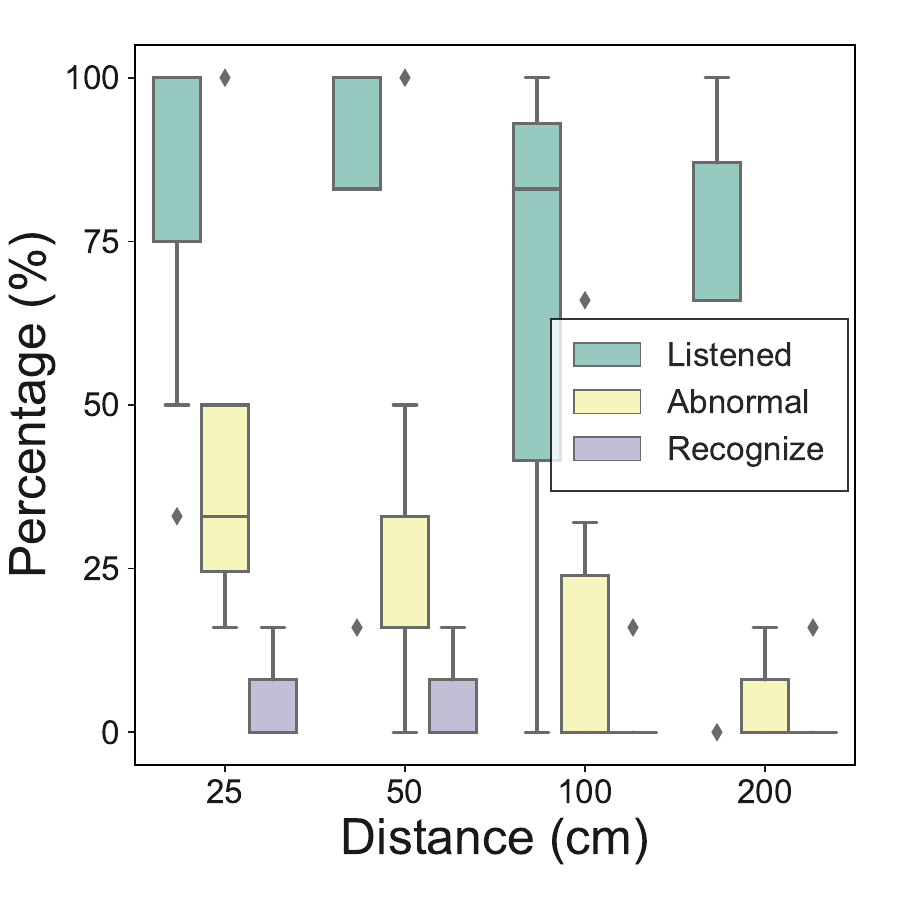}}
\vspace{-10pt}
\caption{Real-world user study of \ours.}
\vspace{-10pt}
\end{figure}

\section{Discussion}\label{sec-discussion-and-limitation}

\subsection{Low-cost Attack}
Table~\ref{tb:cost} lists the cost comparison between PhantomSound and the existing work~\cite{chen2020devil}. The first row records the pricing information of the commercial APIs, which is measured by the duration of given audios (in minutes). 
The recent black-box attack~\cite{chen2020devil} is reported to incur the cost of  1,500 queries for building the substitute models, and every query uses an audio with 25 seconds long. In total, such an attack requires $1500*25/60=625$ minutes to train a surrogate model, and can only generate 10 pre-selected commands. To generate extra commands, the attacker needs to submit additional queries ($\sim$100) for the candidate AEs. Suppose the length of candidate AEs is 6 seconds, the total time cost for generating extra AEs is $6*100/60=10$ minutes. All together, the duration of queried audio is 72.5 minutes for producing one single AE. 
In contrast, \ours does not require a substitute model, and as such, it
only takes $\sim$300 queries and $\sim$2,000 queries of one-second audios to craft an untargeted AE (Ours-U) and a targeted AE (Ours-T) respectively. 
We then present the cost to generate one AE based on the pricing and the query audio length (shown in row 4 and 5).
In the end, \ours saves $93.1\%$ and $65.5\%$ of the cost for crafting an AE, a drastic improvement. 

\subsection{Limitations}

The limitation of \ours includes that: 1) the attack is sensitive to ambient noise; 2) there is no guarantee to generate an AE for any input and any target; 3) this attack could not substantially modify very long sentences; 4) the attack distance is relatively short as presented in Section~\ref{subsec:impact}. To address the first and the forth limitation, the adversary can either amplify the perturbation power or attack the victim in a relatively quiet place. The second and third limitations are possibly addressed using multiple repeated attempts of phoneme injections, which will increase the likelihood of generating a successful perturbation with a potential caveat of growing costs.
\vspace{-10pt}
\begin{table}[h]
\caption{Cost comparison}
\vspace{-10pt}
\centering
\scalebox{0.8}{
\begin{tabular}{l|l|l|l|l}
\hline
\hline
                              & Google  & MS & AMZ  & IBM     \\
                              \hline
Pricing/min                     & \$0.024 & \$0.016   & \$0.024 & \$0.01  \\
\hline
Build model~\cite{chen2020devil} & \multicolumn{4}{c}{625 min}             \\
\hline
Craft AE~\cite{chen2020devil}    & \multicolumn{4}{c}{10 min}              \\
\hline
Total time/AE~\cite{chen2020devil}            & \multicolumn{4}{c}{72.5 min}            \\
\hline
Total time/AE (\textbf{Ours})               & \multicolumn{4}{c}{5 min - 25 min }              \\
\hline
Cost/AE~\cite{chen2020devil}    & \$1.74  & \$1.16    & \$1.74  & \$0.725 \\
\hline
Cost/AE (\textbf{Ours-U})            & \$0.12  & \$0.08    & \$0.12  & \$0.05 \\
\hline
Cost/AE (\textbf{Ours-T})            & \$0.6  & \$0.4    & \$0.6  & \$0.25 \\
\hline
\textbf{Saving/AE} (\textbf{Ours})           & \multicolumn{4}{c}{\textbf{93.1\%}/\textbf{65.5\%}}\\

\hline
\hline
\end{tabular}
}
\label{tb:cost}
\end{table}

\subsection{Defense}
Prior studies~\cite{yang2018characterizing, yuan2018commandersong, li2020advpulse} reveal that the audio adversarial attack can be defended by signal processing techniques,
since the adversarial perturbations are delicately crafted and hence are deemed fragile. The signal processing techniques, however, can reduce the fidelity of perturbations and hence protecting the ASR models. Typical signal processing defense methods include 1) \emph{Down sampling (DS)}: decreasing the sampling rate of AEs to disrupt the quality of AEs~\cite{yuan2018commandersong, li2020advpulse, yang2018characterizing}; 2) \emph{Quantization}: as the original AEs are encoded by 16-bit values, the quantization technique rounds the 16-bit precise value to its nearest integer multiple of $Q$, where $Q$ represents the quantization level. A higher $Q$ results in a lower precision of AEs, which has been adopted to defend against the attacks~\cite{li2020advpulse, yang2018characterizing}. 
3) \emph{Low pass filtering (LFP)}: the defense can use a Butterworth low-pass filter with different cutoff frequencies to remove the high-frequency components of the perturbations~\cite{li2020advpulse}.

We reproduce the aforementioned three defense methods to test their effectiveness against \ours. Specifically, for \emph{DS} approach, we modify the sampling rate of AEs from 16k to 8k and 4k.
In the \emph{quantization} setting, we follow the existing work~\cite{li2020advpulse} to set $Q$ as 256, 512, and 1,024.
Then, we build a Butterworth low-pass filter with a cutoff frequency of 4kHz, and set the order of the filter as 6. To validate the defense performance comprehensively, we generate \textbf{1,190} AEs from 20 clean audio samples and process them with 6 different defense settings. 

\begin{figure}[h]
\vspace{-15pt}
\centering     
\subfigure[Defense performance of down sampling and low pass filter]{\label{fig:df1}\includegraphics[width=40mm]{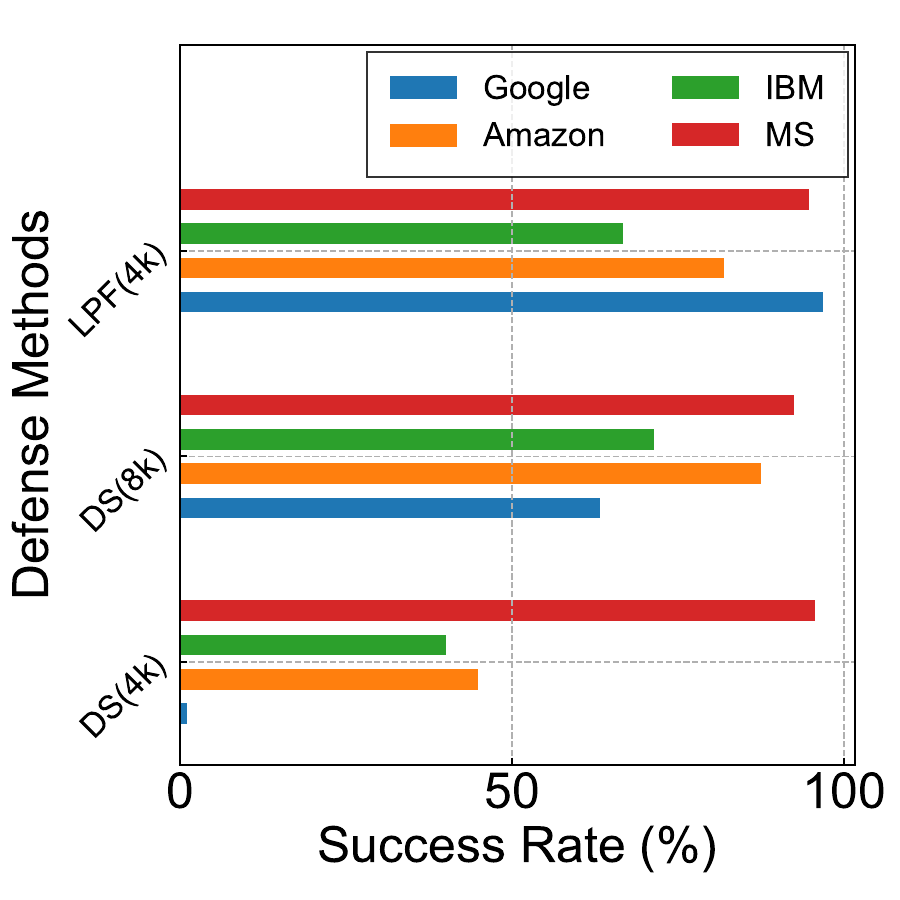}}
\subfigure[Defense performance of quantization]{\label{fig:df2}\includegraphics[width=40mm]{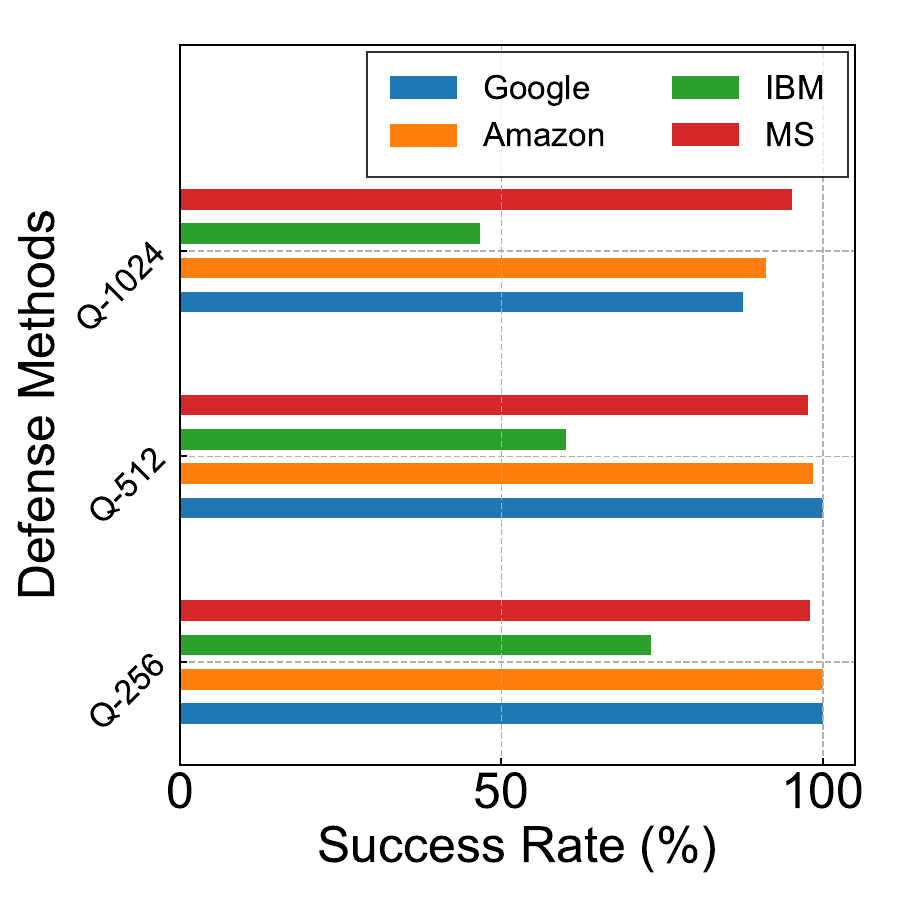}}
\vspace{-10pt}
\caption{Performance of \ours against different defenses. }
\label{fig:defense}
\vspace{-15pt}
\end{figure}

We use the processed AEs to attack 4 commercial ASR APIs, and present the results in Fig.~\ref{fig:defense}. Fig.~\ref{fig:df1} shows that \emph{LPF} can barely impact the attack success rate of AEs and APIs. For comparison, the \emph{DS} technique slightly changes the attack success rate from 100\% to 92.4\% (Microsoft), 71.4\% (IBM), 87.5\% (Amazon), and 63.3\% (Google). This method can further reduce the success rate by applying a lower sampling rate (e.g., with 4k sampling rate, the IBM and Amazon API can defend against $\sim$60\% attacks, while the Google API is not supported for the audio input with such a low sampling rate. Different from the findings from previous work~\cite{li2020advpulse, yang2018characterizing} that quantization is effective in defending against the adversarial attack, our results show a converse performance. From Fig.~\ref{fig:df2}, we observe that only the IBM API can be affected by the quantization, which reduces the success rate to 73\%, 61\%, and 47\% for q=256, 512, and 1,024, respectively. To summarize, our results demonstrate that the existing signal processing-based defense approaches cannot protect the commercial APIs from \ours. Future research on defense mechanisms are needed to provide more secure speech-to-text and voice assistance services.  

\rev{
\subsection{Ethical Issues}
The intention behind publishing this work is to enlighten the academic and tech community about the vulnerabilities of commercial ASR APIs and smart speakers, it may also provide malicious actors with the knowledge and tools to exploit these vulnerabilities for harmful purposes, such as privacy invasion, identity theft, or unauthorized control of connected devices.
If the findings of this paper are misused,
malicious actors could potentially manipulate smart speakers into sharing sensitive information or performing unauthorized actions, there may be potential financial and reputational harm to individuals and corporations. 
To address these ethical concerns, it would be advisable to collaborate with manufacturers of smart speakers to design effective countermeasures to defend against this attack. 
}
\vspace{-5pt}
\section{Related Work}

\label{sec-relate}
The study of adversarial attacks starts from the discovery of intriguing properties of the neural networks around 2014~\cite{goodfellow2014explaining, szegedy2013intriguing}. Researchers manually or automatically add small perturbations to the input and thereby misleading the neural network models. 

\noindent
\textbf{Adversarial Attacks against ASR Systems:}
Existing work~\cite{cisse2017houdini, carlini2018audio, alzantot2018did, qin2019imperceptible} has proposed different optimization algorithms to craft effective AEs towards ASR models with some knowledge of the victim's ASR model (e.g., prediction scores or logits output). However, the robustness of their attack approaches in a real-world over-the-air scenario is usually unverified. 
The recent physical attacks such as CommanderSong~\cite{yuan2018commandersong}, Devil's Whisper~\cite{chen2020devil}, and AdvPulse~\cite{li2020advpulse} require a substantial cost (in time and money) for the attackers to succeed in attacking the black-box voice assistants. 

\noindent \textbf{Signal Processing Attacks: }
Rather than exploiting the vulnerabilities of neural networks in ASR systems, the signal processing attacks aim at attacking the signal pre-processing or feature extraction modules. They usually exploit the discrepancies between the human auditory system and the perceptual hearing system of microphones to fool the ASR system. These attacks analyze the input and output of the feature extraction procedure, and then they modify the input of feature extraction and preserve the shape of output to either hide their attack~\cite{abdullah2019practical} or mislead the ASR system in producing incorrect transcriptions~\cite{abdullah2021hear}. 
Even though the existing signal processing attacks demonstrate the efficiency and effectiveness against the black-box models, it is relatively straightforward to defend against using frequency filters.

\noindent\textbf{Audio Backdoor Attacks:}
Different from adversarial attacks which attack a trained model, backdoor attacks~\cite{liu2018trojaning, gu2019badnets, guo2020practical} inject backdoor triggers during the training process. Recently, researchers demonstrated that the backdoor attack~\cite{shi2022audio, zhai2021backdoor} can also be implemented in the ASR model and Speaker Verification models. To defend against the backdoor attacks in the image domain, several countermeasures are proposed~\cite{guo2023scale, guo2022aeva}.

\noindent \textbf{Other Related Works:}
Some attackers exploit the imperfection of hardware 
(e.g., microphone) to deliver inaudible attacks through different media~\cite{zhang2017dolphinattack, yan2020surfingattack, sugawara2020light, li2023echoattack}. Besides, Danger~\cite{zhang2019dangerous} uses homophones (i.e., different words with similar sounds) to attack ASR skills. Researchers also develop side-channel attack~\cite{wang2022ghosttalk} by injecting voice commands through a power line. Speech synthesis attack produces victim's fake speech by generative models~\cite{wenger2021hello}. To protect the victim's original speech, researchers add perturbations(~\cite{huang2021defending, wang2023vsmask}) to prevent the generating of deep fake speech.

\section{Conclusion}\label{sec-conclusion}

In this work, we proposed \ours, a practical, black-box, and query efficient audio attack against commercial  ASR systems and  IVC devices in a real-world scenario. As opposed to the existing attacks that require prior knowledge of the target model, we propose a phoneme-level searching method to generate AEs and perturbations rapidly and effectively in a black-box setting. In the real-world experiments, \ours is shown to be practical and robust in attacking 5 popular commercial voice controllable devices over the air, which could potentially cause hazard to the smart home.

\vspace{10pt}

\bibliographystyle{plain}
\bibliography{references}
\appendix
\appendix
\section*{Appendix}

\rev{
\section{Liveness Detection Methods}\label{sec-live}

The first liveness detection model~\cite{kinnunen2017asvspoof} (CQCC) is the baseline model of the ASVSpoof challenge, which uses constant-Q cepstral coefficients (CQCC) features and Gaussian Mixture Model (GMM) to separate the natural and
replayed human speech. 
The second detection, STC~\cite{lavrentyeva2017audio} won the ASVSpoof 2017 challenge, which exploits
a Light Convolutional Neural Network (LCNN) to perform detection. The third model called Void~\cite{ahmedvoid} is a fast and high efficient detection algorithm proposed recently. It considers novel spectrogram features such as spectrogram delay patterns, peak patterns, and Linear Prediction Cepstrum coefficient
(LPCC) to achieve state-of-the-art detection accuracy~\footnote{We skip some more advanced liveness detection approaches such as ~\cite{mengyour, li2021robust, guo2022supervoice} because they require extra hardware to facilitate the detection.}.

\section{User Study Setup}\label{sec-ucset}
We recruit 7 volunteers from our institute and 13 volunteers from Amazon Mechanical Turk. Before the experiment, we informed them that their name, voice, and other personal information would not be recorded. We would only release the statistical data about reactions to our attack. For 2 in-person volunteers, we played 6 crafted perturbations at 4 different distances while they were speaking to the smart speaker. For the 5 volunteers from our institute and the 13 volunteers from the MTurk, we asked them to complete the hearing screening before the experiment. Then, we sent them the 6 perturbations and asked them to play the perturbation toward their smart speakers at 4 different distances. 
To ensure that all the results are valid, we verified that there were no random responses (e.g., ``listened" at a far distance but not at a close distance; ``recognized" but not ``listened"). Every experiment took $\sim$10 minutes because some subjects reported that they would need to play perturbations multiple times before determining an answer. }


\end{document}